\documentclass[apj]{emulateapj}
\usepackage[dvips]{color}

\newcommand*{\teff}{$T_{\rm eff}$}

\newcommand*{\kms}{km s$^{-1}$}
\newcommand*{\zmax}{$Z_{\rm max}$}
\newcommand*{\rapo}{$r_{\rm apo}$}
\newcommand*{\Rapo}{$R_{\rm apo}$}
\newcommand*{\rperi}{$r_{\rm peri}$}
\newcommand*{\vphi}{$V_{\rm \phi}$}
\newcommand*{\msun}{$M_\odot$}

\usepackage[dvips]{color}
\usepackage{apjfonts}
\usepackage{graphicx}
\usepackage{natbib}
\usepackage{longtable}

\slugcomment{}

\shorttitle{CEMP Sub-Classes in the Halo System}
\shortauthors{Carollo et al.}

\begin{document}


\title{Carbon-Enhanced Metal-Poor Stars: CEMP-s and CEMP-no Sub-Classes in the Halo System of the Milky Way}

\author{Daniela Carollo\altaffilmark{1}}
\affil{Dept. of Physics and Astronomy - Astronomy, Astrophysics and Astrophotonic Research Center\\
Macquarie University - North Ryde, 2109, NSW, Australia}
\email{daniela.carollo@mq.edu.au}

\author{Ken Freeman}
\affil{Research School of Astronomy \& Astrophysics, Australian National
University\\ \& Mount Stromlo Observatory, Cotter Road, Weston, ACT,
2611, Australia\\}
\email{kcf@mso.anu.edu.au}

\author{Timothy C. Beers}
\affil{National Optical Astronomy Observatory, Tucson, AZ, 85719, USA\\
and JINA: Joint Institute for Nuclear Astrophysics}
\email{beers@noao.edu}

\author{Vinicius M. Placco}
\affil{Gemini Observatory, Hilo, HI 96720, USA}
\email{vplacco@gemini.edu}

\author{Jason Tumlinson}
\affil{Space Telescope Science Institute, Baltimore, MD, 21218, USA}
\email{tumlinson@stsci.edu}

\author{Sarah L. Martell}
\affil{Australian Astronomical Observatory, North Ryde, 2109, NSW, Australia}
\email{smartell@aao.gov.au}

\altaffiltext{1} {INAF-Osservatorio Astronomico di Torino, Pino Torinese, Italy}

\begin{abstract}

We explore the kinematics and orbital properties of a sample of
323 very metal-poor stars in the halo system of the Milky Way,
selected from the high-resolution spectroscopic follow-up
studies of Aoki et al. and Yong et al. The combined sample
contains a significant fraction of carbon-enhanced metal-poor (CEMP)
stars (22\% or 29\% , depending on whether a strict or relaxed criterion
is applied for this definition). Barium abundances (or upper limits) are
available for the great majority of the CEMP stars, allowing for their
separation into the CEMP-$s$ and CEMP-no sub-classes. A new method to
assign membership to the inner- and outer-halo populations of the Milky
Way is developed, making use of the integrals of motion, and applied to
determine the relative fractions of CEMP stars in these two sub-classes
for each halo component. Although limited by small-number statistics,
the data suggest that the inner halo of the Milky Way exhibits a
somewhat higher relative number of CEMP-$s$ stars than CEMP-no stars
(57\% vs. 43\%), while the outer halo possesses a clearly higher
fraction of CEMP-no stars than CEMP-$s$ stars (70\% vs. 30\%). Although larger
samples of CEMP stars with known Ba abundances are required, this result
suggests that the dominant progenitors of CEMP stars in the two halo
components were different; massive stars for the outer halo, and
intermediate-mass stars in the case of the inner halo.


\end{abstract}


\section{Introduction}

The halo system of the Milky Way hosts the vast majority of the
presently observed metal-poor (MP: [Fe/H] $< -$1.0) and very metal-poor
(VMP; [Fe/H] $< -$2.0) stars. These objects are of special interest, as
they preserve a fossil record of the nucleosynthesis products of the
first generations of stars that formed shortly after the Big Bang, and
hence offer a unique opportunity to explore the early stages of galaxy
formation and the chemical evolution of galactic halos.

\subsection{The Halo System of the Milky Way}

Although once thought of as a comparatively simple component, the halo
system of the Milky Way is apparently complex. It comprises at
least two main structural components, the diffuse inner- and the
outer-halo populations, numerous detected sub-structures, such as
streams and overdensities (Grillmair 2009, and references therein), as
well as a substantial number of globular clusters. Recently, ages
determined from white dwarf cooling sequences, and other methods, have
indicated differences (of up to 2 Gyrs) in the ages of clusters and
field stars that can be associated with the inner- and outer-halo
populations \citep{Kalirai2012,Hansen2013a}.

The dual nature of the diffuse halo was initially recognized by
\citet[][hereafter C07 and C10]{Carollo2007,Carollo2010}, based on
analyses of the kinematics of a local sample of calibration stars from
the Sloan Digital Sky Survey \citep[SDSS;][]{York2000,Gunn2006}. These
authors demonstrated that the inner and outer halos possess different
peak metallicities (inner: [Fe/H] $\sim -$1.6; outer: [Fe/H] $\sim
-$2.2), as well as different spatial distributions, with the inner halo
exhibiting a flatter density profile than the nearly spherical outer
halo. According to their analysis, the inner-halo component is
essentially non-rotating, while the outer-halo component exhibits a
significant retrograde signature. The velocity ellipsoids of the
populations also differ, in the sense that the outer-halo population is
clearly ``hotter'' than the inner-halo population.


The proposed duality of the Galactic halo has been confirmed by other
authors, based on different data sets. For example, \citet{deJong2010}
analysed stellar number-density maps constructed from the ``vertical''
photometric stripes obtained during the SEGUE-1 sub-survey of
SDSS (Yanny et al. 2009), and identified a shift in the peak stellar metallicity as a
function of Galactocentric distance ([Fe/H] $\sim -1.6$ at $r <$ 15
kpc ; [Fe/H] $\sim -$2.2 at $r >$ 15-20 kpc), similar to the shift in
the peak of the MDF argued to be present by C07 and C10, and occuring at
the same halo transition radius, $r_{\rm HTR},$\footnote{The halo
transition radius is defined by Tissera et al. (2013b) to
denote the region in their simulated galaxies marking the transition
from relative dominance by the stellar inner-halo population to the
stellar outer-halo population.} as inferred from their kinematical analysis.
\citet{Kinman2012} identified signatures of the retrograde nature of
the outer-halo population from samples of RR Lyrae stars chosen without
kinematic bias, reporting that the transition from a flattened,
essentially non-rotating inner halo to a spherical, retrograde outer
halo occurs beyond about 12.5 kpc. Similar evidence for retrograde
motions have been described in \citet{Hattori2013} and
\citet{Kafle2013}, based on samples of blue horizonal-branch (BHB) stars
with available metallicities and radial velocities from SDSS. Additional
lines of evidence supporting the dual halo of the Milky Way have been
reported by \citet{Beers2012}. Most recently, \citet{An2013} have used
photometric estimates of stellar metallicity for stars in SDSS Stripe 82
to argue that, even in the relatively nearby volume (5-8 kpc from the
Sun), the observed metallicity distribution function (MDF), coupled with
the kinematics of the local halo, is incompatible with a single
population of stars.


The bimodality of the MDF in the halo system of M~31 is now also
well-established. \citet{Sarajedini2012} report evidence of a bimodal
MDF in the halo of M31 with peaks at [Fe/H] $\sim$ $-$1.3 and [Fe/H]
$\sim$ $-$1.9. \citet{Ibata2013} found that the metallicity of the
smooth halo of M~31 decreases from [Fe/H] $\sim$ $-$0.9 at $r = 27$ kpc,
down to [Fe/H] $\sim$ $-$1.9 at $r = 150$ kpc. A similar metallicity
shift is reported by \citet{Gilbert2013} and Gilbert et al. (in
preparation), who identify a clear change in the peak in the MDF of
M~31's diffuse halo population at around $r_{\rm HTR}$ $\sim$ 40 kpc,
from a peak of [Fe/H] $\sim -$0.4 inside this radius to [Fe/H] $\sim
-$1.5 outside this radius. Numerical simulations of hierarchical galaxy
formation (in particular those including baryons; Tissera et al. 2013b,
and references therein) produce dual-halo systems that share many of the
observed properties of the Milky Way and M~31. For discussion of a contrary
view, see, e.g., Sch\"{o}nrich et al. (2011) and Beers et al. (2012).

\subsection{Carbon-Enhanced Metal-Poor Stars in the Halo System of
the Milky Way}

The distribution of carbon on iron ([C/Fe]) for halo stars
provides a clear indication of differences in the chemical nature of the
inner- and outer-halo populations. Carbon-enhanced metal-poor (CEMP)
stars have been defined as a subset of MP and VMP stars that exhibit
elevated [C/Fe] ([C/Fe] $\geq +$1.0; Beers \& Christlieb 2005). Other
authors have used slightly different criteria, e.g., [C/Fe] $ \geq +$0.7
\citep{Aoki2007,Carollo2012,Norris2013b}. In the past two decades, it
has been recognized that $\sim$ 15-20\% of stars with [Fe/H] $< -$2.0
are carbon enhanced \citep{Beers1992,Norris1997,Rossi1999,
Beers2005, Cohen2005, Marsteller2005,Rossi2005,Frebel2006,Lucatello2006,
Norris2007, Carollo2012,Cohen2013,Norris2013b,Spite2013}. This fraction rises to
30\% for [Fe/H] $< -$3.0, to 40\% for [Fe/H] $< -$3.5, and $\sim $75\%
for [Fe/H] $< -$4.0 (with the inclusion an apparently carbon-normal star
with [Fe/H] = $-$4.9 by Caffau et al. 2011). This increasing trend
of CEMP-star frequency with declining [Fe/H] is again confirmed from the
many thousands of CEMP stars found among the several hundred thousand
stars with available medium-resolution spectra from SDSS, SEGUE-1,
and SEGUE-2 (Rockosi et al., in preparation), as described by
\citet{lee2013}.  It is presumably no coincidence that the most
extreme iron-deficient star known, the septa metal-poor ([Fe/H] $<
-7.0$) star SMSS~J031300.36-670839.3, exhibits both carbon enhancement
and the characteristic light-element enhancement associated with the
CEMP-no class \citep{Keller2014}.

\citet{Carollo2012} estimated [C/Fe] for the sample of over 30,000
SDSS/SEGUE calibration stars used by C10. In addition to the rise in the
cumulative frequency of CEMP stars at lower metallicity, they also
confirmed a significant increase in the level of carbon enrichment in CEMP
stars, as a function of decreasing metallicity, growing from
$\langle$[C/Fe]$\rangle$ $\sim +$1.0 at [Fe/H] = $-$1.5 to
$\langle$[C/Fe]$\rangle$ $\sim +$1.7 at [Fe/H] = $-$2.7. In the same
paper it was shown that the carbon-on-iron distribution function (CarDF)
changes dramatically with increasing distance from the Galactic plane.
For $|$Z$|$ $< 5$ kpc, relatively few CEMP stars are identified (over
all metallicities). For distances $|$Z$|$ $> 5$ kpc, the CarDF exhibits
a strong tail towards higher values, up to [C/Fe] $>$ +3.0.
\citet{Carollo2012} sought to establish whether these effects are
related to metallicity, or to changes in the nature of the stellar
population with distance above the plane. In the metallicity interval
$-$2.5 $<$ [Fe/H] $\le$ $-$2.0, there is evidence for a significant
contrast in the frequency of CEMP stars stars that are kinematially
assigned to the inner- and outer-halo components, such that
F$_{CEMP_{out}}$ $\sim$ 2$\times$F$_{CEMP_{in}}$. The increase in the
frequency of CEMP stars with distance from the Galactic plane can then
be understood in the context of the greater fraction of outer-halo
population stars that are found farther from the plane.

In any case, the large fraction of CEMP stars at low metallicity
indicates that significant amounts of carbon were produced in the early
stages of chemical evolution in the universe. The observed contrast of
the CEMP-star fractions in the inner- and outer-halo populations
strengthen the picture that these halo components had different origins.
\citet{Carollo2012} speculated that this could be understood if the
progenitors of stars in the outer-halo population included additional
early-universe sources of carbon production, an idea we explore further
in the present paper.

\subsubsection{CEMP Sub-Classes}

Beers \& Christlieb (2005) specify a nomenclature that identifies a
number of sub-classes for CEMP stars. The CEMP-$s$ stars exhibit
over-abundances of elements produced by the so-called slow
neutron-capture process, or $s-$process, such as barium. These stars are
the most commonly observed sub-class of CEMP stars; around 80\% of
CEMP stars exhibit $s$-process-element enhancements \citep{Aoki2007},
including both the CEMP-$s$ and CEMP-$r/s$ sub-classes (the latter
sub-class indicates stars for which the presence of both $r$- and
$s$-process element enhancements are found). The favored scenario for
the production of CEMP-$s$ (CEMP-$r/s$) stars is mass transfer of
carbon- and $s$-process-enhanced material from the envelope of an
asymptotic giant-branch (AGB) star to its (presently observed) binary
companion \citep[e.g.,][]{Herwig2005,Sneden2008}. Observational evidence
now exists to suggest that the CEMP-$r/s$ stars (and other
$r$-process-element rich stars) were enhanced in $r$-process elements in
their natal gas clouds by previous generations of supernovae (SNe), and
did not require a contribution of $r$-process elements from a binary
companion (see Hansen et al. 2011, 2013b).

By contrast, the CEMP-no stars exhibit no strong neutron-capture-element
enhancements, and are preferentially found at lower metallicity than the
CEMP-$s$ and CEMP-$r/s$ stars \citep{Aoki2007}. Possible progenitors for
CEMP-no stars include massive, rapidly rotating, mega metal-poor (MMP;
[Fe/H] $< -6.0$) stars, which models suggest have greatly enhanced
abundances of CNO due to distinctive internal burning and mixing
episodes, followed by strong mass loss \citep{Hirschi2006,Meynet2006,
Meynet2010}. Another proposed mechanism is pollution of the interstellar
medium by so-called faint supernovae associated with the first
generations of stars, which experience extensive mixing and fallback
during their explosions \citep{Umeda2003,Umeda2005,Tominaga2007,
Tominaga2013,Ito2009,Ito2013,Kobayashi2011,Nomoto2013}.


Extant radial-velocity data support the expected differences in the
binary nature of CEMP-$s$ (CEMP-$r/s$) and CEMP-no stars.
\citet{Lucatello2005} argued that multiple-epoch radial-velocity
measurements of CEMP-$s$ stars are consistent with essentially all
CEMP-$s$ (CEMP-$r/s$) stars being members of binary systems. Although
more data is desired for CEMP-no stars, \citet{Hansen2013b} report that
the fraction of binaries among stars within this sub-class is no higher
than expected for random samples of VMP giants. \citet{Cohen2013} and
\citet{Norris2013b} reach similar conclusions, based on their limited
radial-velocity data for CEMP-no stars.

In this paper we make use of the CEMP sub-classes to explore the
chemical differences between the inner- and outer-halo components of the
Galaxy. We employ a combined sample of over 300 VMP and extremely
metal-poor (EMP; [Fe/H] $< -$3.0) stars recently investigated by
\citet{Aoki2013}, \citet{Norris2013a}, \citet{Norris2013b}, and \citet{Yong2013}.

Stars in this sample have available high-quality, high-resolution
spectroscopy, sufficient to enable the classification of the CEMP stars
they contain into the CEMP-$s$ or CEMP-no sub-classes (we do not attempt
to resolve the CEMP-$r/s$ from the CEMP-$s$ sub-class in this paper). We
use their reported radial velocities, distances, and (where available)
proper motions in order to derive full space motions for these stars.
Then, through use of an energy and angular momentum criterion to assign
each CEMP star to either the inner-halo or outer-halo population, we
evaluate the relative numbers of CEMP-$s$ and CEMP-no stars that
are likely to be associated with each halo component.

The paper is organized as follows. Section 2 summarizes our combined
stellar sample, and describes the radial velocities, distance
estimates, assembly of proper motions, and assignment
of the CEMP sub-classes. In Section 3, the space motions and orbital
properties of the stars in our combined sample are derived, together
with the integrals of motion. This section also
describes our methods for assignment of the stars' membership into the
inner- or outer-halo populations, and reports the relative
numbers of CEMP-$s$ and CEMP-no stars for the two halo components.
Finally, Section 4 considers the implications of these results for the
formation and evolution of the Galactic halo populations.

\section{Description of the Combined Sample}

Our dataset is assembled from the samples of stars discussed in
\citet{Aoki2013}, and in the series of papers by \citet{Norris2013a},
\citet{Norris2013b}, and \citet{Yong2013}. For convenience, we simply refer
to the first source as the Aoki et al. sample, and the latter three
sources as the Yong et al. sample. These studies are based on
high-resolution follow-up observations of VMP and EMP targets identified
in the massive SDSS-I and SDSS-II/SEGUE moderate-resolution
spectroscopic data sets, the Hamburg/ESO (HES) objective-prism survey
\citep{Wisotzki1996,Christlieb2001, Christlieb2003,Christlieb2008}, the
HK survey \citep{Beers1985, Beers1992}, and on additional data compiled
from the literature. Elemental abundances have been adopted from
Cohen et al. (2013) for three stars in the Yong et al. sample, and from
Roederer et al. (2014) for two stars in the Yong et al. sample;
these were adopted when either the Yong et al. sample reported no C or
Ba abundance, or only had an upper limit for one or both of these
elements available. Although clear selection biases for the inclusion of
very metal-poor stars exist in the Aoki et al. and Yong et al. samples
(by design), stars with or without carbon enhancement were not
specifically targeted by either group. As with most all high-dispersion
spectroscopic follow-up studies, preference was given in the selection
of targets to include brighter stars (although in the case of the Aoki
et al. sample, this is curtailed by the bright limit of $g \sim 14.5$
among SDSS stars). However, as emphasized by the An et al. (2013)
analysis of an {\it unbiased} sample of stars (with respect to
metallicity), the fraction of outer-halo stars {\it even in the local
neighborhood} is roughly similar to the fraction of inner-halo stars at
low metallicity (see their Figure 18). Hence, we expect that at least a
representative sample of objects drawn from both populations can be
achieved in the metallicity range explored by the combined sample.

The Aoki et al. sample comprises 137 stars, 28 of which have reported C
determinations, and 42 of which have Ba abundances reported. These stars
were observed at high spectral resolution ($R \sim 30000$), in the course
of four observing runs between March and October 2008, using the High
Dispersion Spectrograph \citep{noguchi2002} at the Subaru Telescope.
This sample includes a large number of warmer main-sequence turnoff
stars, which (at the signal-to-noise obtained for the majority of the
sample, $S/N \sim 20-30$) present a challenge to obtain these
abundance measurements. We also include 190 stars from the Yong et al. sample --
the 38 stars from their ``program sample'', and 152 stars in their
literature compilation. This sample comprises 171 stars with C
measurements or upper limits, and 175 stars with Ba abundances or upper
limits. High-resolution spectra ($22000 < R < 85000$) of the program
sample stars were taken between June 2007 and September 2008, using the MIKE
spectrograph \citep{Bernstein2003} on the Magellan Clay Telescope at Las
Campanas Observatory, the HIRES spectrograph
\citep{vogt1994} at Keck Observatory, or the UVES spectrograph
\citep{dekker2000} on VLT UT2 (Kueyen) at the European Southen
Observatory. The selection of the program stars and the literature
compilation is explained in \citet{Yong2013}, and the determination of
stellar parameters is described by \citet{Norris2013a} and
\citet{Yong2013}. In the case of the Yong et al. literature sample, the
authors adopt the measured equivalent widths from the literature, and
performed their own analysis to determine stellar parameters and
abundances.  Tables 1 and 2 list the observed propoerties of the
Aoki et al. and Yong et al. samples, respectively.

There are four stars in common between the Aoki et al. sample and the
Yong et al. sample, indicated with superscript ``f'' next to the star
name in column (1) of Table 1 and ``j'' in Table 2. For the purpose of
the analysis described below, we adopt the parameters from the Yong et
al., as they were able to obtain estimates or upper limits on [C/Fe] and
[Ba/Fe] for these four stars, while \citet{Aoki2013} (using lower $S/N$
spectra) reported no measurements of these abundance ratios. There are
thus 323 unique stars in our combined sample. This sample is pared
further, as described below, to only include stars for which kinematical
parameters could be estimated.

\subsection{Radial Velocities}

Radial velocities for the stars in the Aoki et al. sample are taken from
their high-resolution spectroscopic determinations, which have errors
typically less than 1 \kms. Our adopted radial velocities are listed in
column (5) of Table 1, with the associated errors given in column (6).

Radial velocities for the stars from the Yong et al. sample are taken from
their high-resolution spectroscopic determinations, which have errors
typically less than 1 \kms, although some have errors as large as 3
\kms.  Our adopted radial velocities and errors for this sample are listed in
columns (5) and (6) of Table 2, respectively.

\subsection{Distance Estimates}

Distances for the stars in the Aoki et al. sample were assigned making
use of their surface gravity estimates in order to classify stars as
dwarfs, main-sequence turnoff stars, or giants, then using the $V$
magnitudes and adopted $E(B-V)$ reddening provided by Aoki et al.
(2013) to apply the procedures described by \citet{Beers2012}, and
references therein. These procedures are based on a set of absolute
magnitude relationships (using absorption and reddening-corrected
Johnson V magnitudes and B-V colors) calibrated to Galactic globular and
open clusters, as described by Beers et al. (2000; their Table 2). As
demonstrated in \citet{Beers2000}, photometric distances estimated for
their sample are in good agreement with distances derived from accurate
Hipparcos parallaxes. We have taken care to apply the revised procedure
described by \citet{Beers2012} in carrying out the luminosity-class
assignments prior to assigning distances. The resulting distances should
be accurate to on the order of 10\%-20\%, based on previous tests. Our
adopted distances are listed in column (13) of Table 1.

\citet{Norris2013b} derived distances for their stars via a
straightforward spectroscopic parallax method -- using the final stellar
parameters from \citet{Yong2013}, they find the absolute magnitude ${\rm
M}_{V}$ predicted by the ${\rm Y}^{\rm 2}$ isochrones for each star,
then use the available apparent $V$ magnitudes and {\rm E(B-V)} reddening
values to calculate distances. Note that \citet{Norris2013b} only published
distances for stars in their sample with [FeH] $\le -3.1$; J. Norris
(private communication) kindly applied this approach for the remaining
stars in the Yong et al. sample for our use. Our adopted distances for these
stars, which are expected to have similar accuracy as the distances for
the \citet{Aoki2013} sample, $\sim$ 15\%, are listed in column (13) of
Table 2.

There are two stars in the Yong et al. sample for which well-measured
distance estimates are not available.  These stars, indicated with an ``a''
superscript next to the star name in column (1) in Table 2, are dropped
from further consideration in the kinematic analysis that follows.

\subsection{Proper Motions}

Proper motions for the Aoki et al. stars were taken from the SDSS
stellar database, if the measurement satisfied the criteria of
\cite{Munn2004}, designed to eliminate spurious reported motions. Note
that all proper motions have been corrected for the systematic error
described by \citet{Munn2008}. The adopted proper motions are listed in
columns (9) and (10) of Table 1. The typical errors for individual
proper-motion components (listed in columns (11) and (12) of Table 1)
are on the order of $\sim$ 2-3 mas~yr$^{-1}$.

There are two stars in the Aoki et al. sample for which well-measured
proper motions are not available. These stars, indicated with a ``b''
superscript next to the star name in column (1) in Table 1, are dropped
from further consideration in the kinematic analysis that follows.

Proper motions for the Yong et al. stars have been obtained by matching
the sample stars with the UCAC-4
\citep{Zacharias2013}, SPM-4 \citep{Girard2011}, or PPMXL \citep{Roeser2010}
catalogs. Our adopted proper motions are listed in columns (9) and (10)
of Table 2. The typical errors for individual proper-motion components
(listed in columns (11) and (12) of Table 2) are on the order of $\sim$
2-4 mas~yr$^{-1}$, although a number of stars have substantially higher
proper-motion errors.

\subsection{CEMP Classifications}

Carbon on iron, [C/Fe], and barium on iron, [Ba/Fe], abundance
ratios are reported for the Aoki et al. and Yong et al. samples
in columns (14) and (15) of Tables 1 and 2, respectively.
We have assigned a set of codes to each star in the combined sample,
given in columns (16) and (17) of Tables 1 and 2, to indicate the
availability and status of a C and Ba measurement for that star. In
column (16), C = 0 indicates the absence of a C measurement; C = 1
indicates that the carbon status for that star is indeterminate (an
upper limit on [C/Fe] $ > +0.7$); C = 2 indicates that the star is
considered C-normal ([C/Fe] $< +0.7$); C = 3 indicates that the
star is considered a CEMP star ([C/Fe] $\ge +0.7$.) In column (17), Ba =
0 indicates the lack of a Ba measurement; Ba = 1 indicates that the Ba
status for that star is indeterminate (an upper limit on [Ba/Fe] $ > 0.0$);
Ba = 2 indicates that the star is considered Ba-normal (0.0 $<$
[Ba/Fe] $< +1.0$); Ba = 3 indicates that the star is considered
Ba-enhanced ([Ba/Fe] $\ge +1.0$); Ba = 4 indicates that the star is
considered Ba-deficient ([Ba/Fe] $\le 0.0$)

Note that many of the stars in the combined sample lack Eu
measurements, which precludes use of the [Ba/Eu] abundance ratio
necessary to discriminate between the CEMP-$s$ and CEMP-$r/s$
sub-classes, according the taxonomy of \citet{Beers2005}. However, as
noted above, our current understanding of the progenitors of the
CEMP-$r/s$ sub-class necessarily involves the same mass-transfer episodes as
for the CEMP-$s$ stars, presumably onto a star that had formed from an
ISM that had already been enhanced in $r$-process elements. Note that no
Eu abundance measurement is required in order to assign the CEMP-no
sub-class.

For the purpose of our following analysis, we classify the CEMP stars
with [Ba/Fe] $\ge +1.0$ as CEMP-$s$ (which could include some CEMP-$r/s$
stars); those with [Ba/Fe] $\le 0.0$ are considered CEMP-no stars.
Column (18) of Tables 1 and 2 list the assigned sub-classifications for
the CEMP stars in the Aoki et al. and Yong et al. samples,
respectively. There are a total of 70 stars in the combined sample
which are classified as CEMP, of which 40 are assigned to be CEMP-$s$,
24 are assigned to be CEMP-no, and 6 are not sub-classifiable, due to a
missing Ba measurement. We refer to these stars as our ``Strict Sample''
(hereafter, SS), indicating that the stars passed all tests needed to
clearly assign their status. There are two additional stars that we add
to the SS, the canonical CEMP-no stars HE~0107-5240
\citep{Christlieb2002} and HE~1327-2326 \citep{Frebel2005}, both of which
are formally considered Ba-indeterminate, due to their high upper limits
on [Ba/Fe], but that clearly belong to this sub-class on the basis of their
light-element abudance patterns. The 6 stars that were not sub-classifiable,
but which meet the strict criterion for inclusion as CEMP stars, we label
``CSS,'' indicating that they are carbon-enhanced stars.

Because we are battling low-number statistics, we have also considered
another sub-sample of stars, which we refer to as the ``Extended Sample''
(ES). This sample includes all of the SS (and CSS) stars, as well as (1) stars
that are within reasonable error bars (as summarized in the table notes)
of being classified as CEMP, and/or CEMP-$s$ and CEMP-no, (2) stars that
have light-element abundance patterns that suggest a CEMP-no
classification, (3) stars that are expected to be CEMP-no, based on their low
metallicity and the detection limit for Ba at their (lower) temperatures
in the Aoki et al. sample, or (4) stars that are considered CEMP
according to the luminosity criterion of \citet{Aoki2007}\footnote{The
\citet{Aoki2007} luminosity criterion is: [C/Fe] $\ge +0.7$, for
log($L/L_\odot$) $\le 2.3$ and (ii) [C/Fe] $\ge +3.0 -
\log(L/L_\odot$), for log($L/L_\odot$) $> 2.3$.}.
Such stars are indicated by superscripts ``d'' or ``e'' next to the star
names in column (1) of Tables 1 and 2.

There are a total of 96 stars in the combined sample that meet the more
relaxed criteria described above, of which 42 are assigned to be
CEMP-$s$, 46 are assigned to be CEMP-no, and 8 stars that are not further
sub-classifiable, due to a missing Ba measurement; these 8 stars we label
``CSS'' or ``CES,'' indicating that they are carbon-enhanced stars. By
definition, the ES includes all of the stars in the SS, and the CES
stars include the CSS stars. Column (19) of Tables 1 and 2
provides a four digit code, where each digit takes on the value 1 or 0,
depending on its classification according to this scheme. The digits
correspond to the classes SS, ES, CSS, or CES. For example, a code
1100 would indicate that the star is a member of the SS and the ES,
but not the CSS or the CES.

The upper panel of Figure 1 shows the distribution of [Fe/H] for all the
stars in our combined sample. In the bottom panel the unshaded histogram
represents the MDF for the 88 CEMP stars of the SS or ES (as well as the 8
CSS or CES stars, which could not be further sub-classified), the light
gray distribution applies to the 42 SS or ES stars assigned to the CEMP-$s$
sub-class, and the cross-hatched distribution indicates the MDF for the
46 SS or ES stars assigned to the CEMP-no sub-class. All of the CEMP-no stars
are located in the metallicity regime [Fe/H] $< -$2.5, in agreement with
the metallicity trend first identified by \citet{Aoki2007}. The lowest
metallicity CEMP$-s$ star in our sample has [Fe/H] = $-$3.5.

\section{Kinematic Analysis}

\subsection{Derivation of Space Motions and Orbital Properties}

The combination of radial velocities, distances, and proper motions for
our sample stars are used to calculate the full space motions. The
velocity components ($U, V, W$) are relative to the Local Standard of
Rest (LSR), where $U$ is taken to be positive in the direction toward
the Galactic anticenter, $V$ is positive in the direction of Galactic
rotation, and $W$ is positive toward the North Galactic Pole. The
velocities are corrected for the motion of the Sun with respect the LSR
by adopting the values $(U,V,W)$ = ($-9$,12,7) km~s$^{-1}$ (Mihalas \&
Binney 1981). We have also evaluated the rotational velocity in the
cylindrical Galactocentric reference frame, with its origin at the
center of the Galaxy. This velocity is denoted as V$_{\phi}$, and is
calculated assuming that the LSR is on a circular orbit with a value of
220 km~s$^{-1}$ (Kerr \& Lynden-Bell 1986). In our calculations, we
assume a value for the location of the Sun at R$_{\sun}$ = 8.5 kpc.
These values are consistent with recent independent determinations of
these quantities by \citet{Ghez2008}, \citet{Koposov2009}, and \citet{Bovy2012}.

The orbital parameters of the stars are derived by adopting an analytic
St\"ackel-type gravitational potential, which consists of a flattened,
oblate disk, and a nearly spherical massive dark-matter halo (see the
description given by Chiba \& Beers 2000, Appendix A). The peri-Galactic
distance, $r_{\rm peri}$, is defined as the closest approach of an orbit
to the Galactic center, while the apo-Galactic distance, \rapo, is the
farthest extent of an orbit from the Galactic center. The orbital
eccentricity, $e$, is defined as $e$ = (\rapo\ $-$ \rperi) /(\rapo\ +
\rperi), while \zmax\ is the maximum distance of a stellar orbit above
or below the Galactic plane. In addition, we evaluate the integrals of
motion for any given orbit, deriving the energy, $E$, and
the angular momentum in the vertical direction, $L_{\rm Z}$ =
$R\times$$V_{\phi}$. Note that $R$ represents the distance from the
Galactic center projected onto the disk plane. Typical errors
on the orbital parameters (at $Z_{\rm max}$ $<$ 50 kpc; C10) are:
$\sigma_{r_{peri}}$ $\sim$ 1 kpc, $\sigma_{r_{apo}}$ $\sim$ 2 kpc,
$\sigma_{ecc}$ $\sim$ 0.1, $\sigma_{Z_{max}}$ $\sim$ 1 kpc.

Based on the derived orbital parameters, there are four stars
in the Aoki et al. sample and 14 stars in the Yong et al. sample that
are formally unbound from the Galaxy.  These stars, indicated with a ``c''
superscript next to the star name in column (1) of Tables 1 and 2, are
dropped from further consideration in the kinematic analysis that
follows.

Taking into account all of the stars dropped from the two samples, as
described above, we proceed with the kinematic analysis based on 127
stars from the Aoki et al. sample (4 stars that also appear in the Yong
et al. sample are dropped) and 174 stars from the Yong et al. sample,
for a total of 301 stars. Among the CEMP stars, there are a total of 64
stars suitable for the kinematic analysis in the SS, and 88 in the ES.
In order to avoid confusion, we refer to this full set of program stars
with suitable kinematic determinations as the ``Total Kinematic Sample"
and the sub-sample of CEMP stars within it as the ``CEMP Kinematic
Sample.''

\subsection{The Energy-Angular Momentum Diagram: Membership Assignment
to the Inner- and Outer-Halo Populations}

We now seek to evaluate the relative numbers of CEMP-$s$ and CEMP-no stars in the
inner- and outer-halo populations contained in the CEMP Kinematic
Sample, which requires that a population
assignment be made. In \citet{Carollo2012}, such a division  was accomplished by
employing the extreme deconvolution technique \citep{Bovy2011},
applied to the rotational velocity distribution of the low-metallicity
sub-sample of the SDSS/SEGUE DR7 calibration stars.  This technique is not
suitable for the present sample, due to the limited number of stars that
are available.  We explore a new method below, comparing with a large
external sample of stars, before returning to the kinematic analysis of
our program stars.

\subsubsection{A New Method of Kinematic Population Assignment; Testing the
SDSS/SEGUE DR7 Calibration Stars}

An alternative method to that previously used employs the integrals
of motion, such as the total energy, $E$, and the angular momentum in
the vertical direction, $L_{\rm Z}$. To explore this approach, we have
calculated the integrals of motion for the SDSS/SEGUE DR7 calibration stars
used by C10\footnote{The distance estimates for the stars in this sample
are based on the luminosity-class assignments revised according to the
procedures descibed by \citet{Beers2012}.}. The results are shown in
Figure 2 as an energy-angular momentum diagram, also known as the
Lindblad diagram.

The point sizes in the two panels of Figure 2 are proportional to the
logarithmic number density at a given point of the diagram, while the
colors denote different ranges of metallicity. According to C10, the
thick disk has a mean Galactocentric rotational velocity of
$\langle$\vphi$\rangle$ = 182 km~s$^{-1}$, and a velocity dispersion,
$\sigma$(\vphi) = 51 km~s$^{-1}$, which correspond to a mean angular
momentum of $\langle$$L_{\rm Z}$$\rangle$ $\sim$ 1550 kpc~km~s$^{-1}$,
and a range of angular momentum between $L_{\rm Z}$ $\sim$ 1100 and 2000
kpc~km~s$^{-1}$ (at the one-sigma level).
Thick-disk stars are highly bound to the Galaxy,
have a peak in their MDF at [Fe/H] $\sim -$0.6, and dominate the light
blue area in the Lindblad diagram (upper panel). The metal-weak
thick-disk (MWTD) population, when treated as an independent component
(as in C10), exhibits a mean Galactocentric rotational velocity
$\langle$\vphi$\rangle$ $\sim$ 100-150 kpc~km~s$^{-1}$, and a velocity
dispersion of $\sigma_{V_{\phi}}$ $\sim$ 40 km~s$^{-1}$. These values
correspond to a mean angular momentum of $\langle$$L_{\rm Z}$$\rangle$
$\sim$ 1000 kpc~km~s$^{-1}$, and angular momentum in the range 700-1300
kpc~km~s$^{-1}$ (one-sigma). MWTD stars are highly bound to the
Galaxy, exhibit a peak in their MDF at [Fe/H] $\sim -$1.2 (C10), and are
clearly present in the Lindblad diagram, as
represented by red filled dots in the top panel of Figure 2.

C07 and C10 have shown that the inner-halo population is dominated by
stars on high-eccentricity orbits, with the majority of inner-halo
stars having apo-Galactic distances not greater than $r_{\rm apo}$ = 15
kpc, a mean Galactocentric rotational velocity $\langle$\vphi$\rangle$
$\sim$ 0, with a velocity dispersion of $\sigma_{V_{\phi}}$ $\sim$ 100 km
s$^{-1}$ (roughly a Gaussian distribution symmetrical about $L_{\rm Z}$
$\sim$ 0). Based on these values, the location of the majority of the inner-halo
population in the Lindblad diagram would be in the range of $-850$
kpc~km~s$^{-1}$ $< L_{\rm Z}$ $<$ 850 kpc~km~s$^{-1}$ (at the one-sigma
level), and mostly clustered in a narrow region around $L_{\rm Z}$ = 0
kpc~km~s$^{-1}$. This is clearly visible in the top panel of Figure 2,
where the green area denotes stars with metallicity in the range $-$2.0
$<$ [Fe/H] $\le -$1.3; this interval was selected to bound the metallicity peak of the
inner-halo population ([Fe/H] = $-$1.6), and to reduce
contamination from MWTD stars.

The inner-halo stars are mostly highly bound (lower energy
values), but also comprise stars with higher energy orbits, $ E> -$1.0
km~$^{2}$ s$^{-2}$ (in units of 10$^{5}$), and are clearly visible in
the top panel of Figure 2. The outer-halo population is characterized by
a wide range of eccentricity values, with a significant number of
outer-halo stars having low to intermediate orbital eccentricities, and
the majority having apo-Galactic distances \rapo\ $>$ 15 kpc. The
mean Galactocentric rotational velocity of the outer-halo stars is
$\langle$\vphi$\rangle$ $\sim -$85 km~s$^{-1}$, with a velocity
dispersion of $\sigma_{V_{\phi}}$ = 165 km~s$^{-1}$. In the Lindblad
diagram, the outer-halo stars are primarily located in the region
characterized by retrograde motions, at $L_{\rm Z}$ $< -$ 700
kpc~km~s$^{-1}$, and with high-energy orbits. The metallicity peak of
the outer halo is [Fe/H] $\sim -$2.2, and includes many of the stars
with [Fe/H] $< -$2.0, represented in Figure 2 (top panel) by black
filled circles.


In the bottom panel of Figure 2, the red dashed curve denotes the locus
that would apply for stars possessing orbits with apo-Galactic distance
\rapo\ = 15 kpc. This is derived using the expression for the energy
$E$ = $L_{\rm Z}^2$/2r$^2$ + $\Phi(r,Z)$, and V$_{\rm r}$ = 0, where $\Phi(r,Z)$
is the Galactic potential, and $r$ is in kpc. The locus of the
points in the ($E$, $L_{\rm Z}$) plane for $r_{\rm apo}$ = 15 kpc and
$Z$ = 0 is $E$ = ($L_{\rm Z}^2$)/(2$\times$15$^2$) + $\Phi(15,0)$. All
of the data points located below this curve have orbits with \rapo\ $<$
15 kpc, while those located above the curve have \rapo\ $>$ 15 kpc. The
inner-halo population is characterized by stars with \rapo\ $<$ 15 kpc,
and dominates the distribution below this apo-Galactic radius. Stars
with metallicity [Fe/H] $\le -$1.3 and apo-Galactic radii \rapo\ $<$ 15
kpc are color-coded green in the figure, while stars in same range of
metallicity and apo-Galactic radii \rapo\ $>$ 15 kpc are color-coded
blue. Inspection of this part of the diagram reveals that there are a
significant number of inner-halo stars with apo-Galactic radii exceeding
15 kpc, however they are mainly clustered below a binding energy of $E$
= $-$1.0 km$^{2}~s^{-2}$ (in units of 10$^{5}$; represented by the black
horizontal line). Above this value of total energy, the stars with
\rapo\ $>$ 15 kpc are dominated by retrograde motions.

Another important feature can be noticed from inspection of the bottom
panel of Figure 2. The nature of the distribution changes as the energy
increases (stars are less bound to the Galaxy), from being roughly a
Gaussian distribution centered at $L_{\rm Z}$ $\sim$ 0 kpc~km~s$^{-1}$
and $E$ = $-$1.1 km$^{2}~s^{-2}$, to becoming increasingly retrograde.
The change seems quite abrupt. At the higher energies, the distribution
does not appear Gaussian, and for $E$ $>$ $-$0.6 km$^{2}~s^{-2}$, it
appears more like a uniform distribution. The more bound Gaussian
component could be interpreted as a dynamically relaxed low-$L_{\rm Z}$
system, which is consistent with a component that formed \emph{in-situ}.
The more uniform component could be interpreted as an unrelaxed system
made up of accreted material, reflecting the ($E$, $L_{\rm Z}$)
distribution at the time that the incoming satellites (the presumed
progenitors for many of the outer-halo stars) were disrupted.

The abrupt change of the ($E$, L$_{\rm Z}$) distribution is easily
visualized in Figure 3, where the vertical angular momentum is plotted
against the total energy. The solid line represents a moving average,
with overlapping bins of 200 points. In this figure, only the stars with
\Rapo\ $>$ 15 kpc and [Fe/H] $\le -$1.3 are selected. In this way, we avoid
most of the contamination from the MWTD, but are sure to include most
of the stars that are members of the inner halo. It is worth noting
that, in the region between $E$ $\sim -$1.05 km$^{2}$~$s^{-2}$ and $-$0.9
km$^{2}$~$s^{-2}$ (in units of 10$^{5}$), there is a clear abrupt change
of the mean angular momentum towards retrograde values, going from
$\langle$$L_{\rm Z}$$\rangle$ $\sim$ 0 kpc~km ~s$^{-1}$ to
$\langle$$L_{\rm Z}$$\rangle$ $\sim$ $-$600 kpc~km~s$^{-1}$
(corresponding to $\langle$\vphi$\rangle$ $\sim$ 0 km~s$^{-1}$ and
$\langle$$V_{\phi}$$\rangle$ $\sim$ $-$70 km~s$^{-1}$, respectively).
The angular momentum becomes more retrograde as the energy increases,
achieving a value of $\langle$$L_{\rm Z}$$\rangle$ $\sim$ $-$1400
kpc~km~s$^{-1}$ (corresponding to $\langle$\vphi$\rangle$ $\sim -$165
km~s$^{-1}$). Note that $\langle$\vphi$\rangle$ = $-$70 is very close to
the mean rotational velocity of the outer-halo population
($\langle$\vphi$\rangle$ = $-$80 km~s$^{-1}$), and $\sigma_{V_{\phi}}$
$\sim$ 165 km~s$^{-1}$ corresponds to the velocity dispersion of the
outer halo at the one-sigma level. The abrupt change in the $L_{\rm
Z}-$$E$ diagram shown in Figure 3 can be considered the transition
zone between the inner- and outer-halo populations. The mid-point of
this zone is $E$ $\sim -$ 0.97 km$^{2}$~$s^{-2}$, with an extension up
to $-$ 0.9 km$^{2}$~$s^{-2}$ on the right end, and $-$1.05
km$^{2}$~s$^{-2}$ on the left end. Below $E$ $\sim$ $-1.05$
km$^{2}$~s$^{-2}$, the mean angular momentum averages to $\sim$ 0
kpc~km~s$^{-1}$, while above E $\sim$ $-$0.9 km$^{2}$~s$^{-2}$,
the mean angular momentum is $\sim$ $-$1000 kpc~km~s$^{-1}$
(corresponding to $\langle$\vphi$\rangle$ $\sim -$123 km~s$^{-1}$). For
the purpose of our analysis, we consider an extended transition zone of
$-$1.10 $<$ $L_{\rm Z}$ $< -$0.82 km$^{2}$~s$^{-2}$, in order to take
into account the uncertainties.

Based on these properties, among the stars with apo-Galactic distance
\Rapo\ $> 15$ kpc and [Fe/H] $\le -$1.3, those with binding energy $E <
-$1.1 km$^{2}$~s$^{-2}$ are likely pure inner-halo stars, while those
with binding energy above $E$ $\sim -$0.8 km$^{2}$~s$^{-2}$ are likely
pure outer-halo members. The stars with energy and angular momentum
falling in the transition zone of Figure 3 have similar probability of
being either inner- or outer-halo stars, according to this criterion.
The numbers of stars assigned to the various components from among
the 8039 SDSS/SEGUE DR7 calibration stars shown in Figure 3 are: outer
halo (274), inner halo (6995), and transition zone (770), respectively.
Stars that are on unbound orbits were already dropped from this sample,
so are not represented in the figure. For completeness, we note that if
the metallicity limit is set to [Fe/H] $\le -2.0$, the numbers of stars
assigned to the various components are: outer halo (62), inner halo (1237),
and transition zone (159), respectively.

We now return to our program sample, and apply the above approach for the assignment of stars
into the inner- and outer-halo populations, based on their energies and
intergrals of motion.

\subsection{Relative Numbers of CEMP-s and CEMP-no Stars in the Inner and Outer halo}

The above population assignment procedure is applied to the Total
Kinematic Sample of  Aoki et al. and Yong et al.
Column (20) of Tables 1 and 2 lists the assigned membership for the
stars in these samples, respectively. The label ``O'' indicates outer halo,
``I'' indicates inner halo, ``T'' indicates the transition zone, and
``U'' indicates that the star was unclassified (star is on an unbound
orbit, or is missing the information required for making this
assignment).

The left-hand panel of Figure 4 shows the Lindblad diagram for the stars in
our Total Kinematic Sample (stars in the U class are obviously not
shown). The black filled circles represent C-normal stars,
while the red stars denote the CEMP stars (ES and CES). In the right-hand panel
of Figure 4, the CEMP-$s$ stars having orbits with \rapo\ $<$ 15 kpc are
represented by purple filled triangles, while those with \rapo\ $>$ 15
kpc are denoted by yellow filled diamonds. In the same way, the CEMP-no
stars with \rapo\ $<$ 15 kpc are shown as blue asterisks, while
those with \rapo\ $>$ 15 kpc are represented by light blue filled
squares. The gray dashed curve represents the locus of stars with orbits
having \rapo\ $>$ 15 kpc.

The orbital properties found in the previous subsection suggest
that stars with apo-Galactic distances below 15 kpc can be considered
pure inner-halo members. Stars with apo-Galactic distance greater than
15 kpc and binding energy $E$ $< -$1.1 km$^{2}$~s$^{-2}$ can be
assigned to the inner-halo component as well. Stars with apo-Galactic distance
above 15 kpc and binding energy $E$ $> -0.8$ km$^{2}$~s$^{-2}$ are
likely to be pure outer-halo stars. The remaining stars with \rapo\ $>$
15 kpc and energy between $E$ $\sim -$1.1 km$^{2}$~s$^{-2}$ and
$E$ $\sim -$0.8 km$^{2}$~s$^{-2}$ fall in the transition zone
between the inner-halo population and the outer-halo population, and a
clear membership assignment cannot be made.

Table 3 summarizes the population assignments for stars in the
Total Kinematic Sample and those in the CEMP Kinematic Sample. Note that
for the CEMP stars, the table lists two numbers for each of the
membership assigments, the first corresponding to stars in the SS, the
second (in parenthesis, and always larger) corresponding to stars in the
ES.

For the stars in the CEMP Kinematic Sample assigned to the outer halo,
the SS only contains a total of 12 stars, which is too small to obtain a
meaningful comparison of the CEMP-$s$ and CEMP-no subclasses. Thus, for
our counting exercise, we prefer to make use of the 20 stars in the ES
assigned to the outer halo. In this instance, the counts in Table 3
indicate that 70\% (14/20) of the stars are CEMP-no, and 30\% (6/20) are CEMP-$s$.

For the stars in the CEMP Kinematic Sample assigned to the inner halo,
the 37 stars contained in the SS are sufficient to proceed, and we
prefer to use this sub-sample, due to the more secure classifications. In
this instance, the counts in Table 3 indicate that 43\% (16/37) of the
stars are
CEMP-no, and 57\% are CEMP-$s$. We caution that, if we were to have
compared the proportions of stars in the inner halo based on the ES
members, the fraction of CEMP-no stars would be greater (56\%) than the
CEMP-$s$ stars (44\%), underscoring the need for more, and better,
classifications of CEMP stars for future tests of our hypothesis.

A one-tailed Z-test of the significance of the change in the proportions
of CEMP-no stars in the inner (SS) and outer halo (ES), which
automatically takes into consideration of the relative sample sizes,
rejects the null hypothesis that the fraction is constant ($p = 0.02$).
Even so, this study would clearly benefit from larger samples of
confidently classified CEMP-no and CEMP-$s$ stars. For now, we can only
tentatively suggest that the CEMP-no stars represent the majority of
C-rich stars associated with the outer-halo population, while the
CEMP-$s$ stars represent the majority associated with the inner-halo
population, at least in this sample of very and extremely
low-metallicity CEMP stars.

\newpage
\section{Implications for the Formation of the Halo System}

We find that the relative numbers of CEMP-no stars compared to CEMP-$s$
stars varies between the inner- and outer-halo components of the Milky
Way, with the frequency of CEMP-no stars being higher in the
counter-rotating outer halo, and the frequency of CEMP-$s$ stars being
higher in the non-rotating inner halo. This trend may reflect some
underlying difference in the chemical-enrichment histories of the halo
components, perhaps due to varying yields in the stellar populations
formed by the progenitors of the two components.

The exact origins of the C-enhancement for CEMP stars are not yet
fully understood. In the case of the CEMP-$s$ stars, mass transfer
between binary companions is strongly implied by the abundance patterns
and radial-velocity measurements. \citep{Herwig2005,Sneden2008,
bisterzo2011, bisterzo2012,lugaro2012,placco2013}. This scenario
requires that the binary system form with a low-mass star (the one
observed today) and a star in the right mass range to pass through an
AGB phase, with significant production of carbon and $s$-process
elements, roughly 1.3-6 \msun. Thus, the CEMP-$s$ phenomenon is
associated with intermediate-mass progenitor stars.

The origin of the C-enhancement for the CEMP-no stars is less clear. One
possible scenario involves binary pairings of a low-mass star (the one
observed today) and a more massive binary companion that can donate
carbon in a mass-transfer event, but which did not produce $s$-process
elements \citep{Suda2007}. Another proposed scenario involves ``faint
SNe'', or core-collapse SNe that exploded with low energy, and undergo a
mixing and fallback process, ejecting the C-rich outer layers of the
progenitor star but not material from the inner, Fe-rich core regions
\citep{Umeda2003,Umeda2005,Tominaga2007,Tominaga2013,Ito2009,Ito2013,
Nomoto2013}. In this scenario, CEMP-no stars would arise preferentially
from massive progenitors, roughly 10-60 \msun.
\citet{Hirschi2006}, \citet{Meynet2006}, and \citet{Meynet2010} have
argued that massive, rapidly rotating, MMP stars, with greatly enhanced
abundances of CNO due to distinctive internal burning and mixing
episodes, can blow strong winds that can pollute the primordial ISM with
this material, with the inner regions possibly collapsing into black
holes. Recently, \citet{Chiappini2013} have argued that such sources
may be capable of driving an $s$-process, but its nature is at present
poorly understood. In any event, the rapidly increasing frequency of
CEMP stars with declining [Fe/H] provides clear evidence that sources
capable of producing carbon {\it without} substantial release of
iron-peak elements must be common at early times.

If we assume the binary mass-transfer scenario applies for CEMP-$s$
stars, and the faint SNe and/or rapidly rotating MMP scenario applies
for CEMP-no stars, the implication is that the former dominates in the
stellar populations associated with the inner halo, while the latter are
more numerous in the populations comprising the outer halo. If the
CEMP-$s$ and CEMP-no phenomena vary with the mass of the stars that
produced the carbon, this requires some variation in the IMF for the
progenitor populations of the inner and outer halos. A link between the
CEMP phenomenon and the early-universe IMF was suggested previously by
\citet{Lucatello2005}, \citet{Suda2007}, \citet{Tumlinson2007a},
and \citet{Tumlinson2007b}, among others. The necessary variations can
be accomodated with relatively small changes to the IMF: the ratio of
the number of CEMP-no progenitors (10-60 \msun) to the number of AGB
progenitors (1.3 -- 3.5 \msun) varies from 0.1 for a Salpeter IMF
(power-law slope $\alpha = -2.35$) up to 0.5 for an IMF with a slope
$\alpha = -1.5$. The implication is that the sub-galactic
fragments from which the outer halo was assembled formed stars with a
flatter IMF, richer in massive stars leading to favored production of
CEMP-no stars, while the inner halo was formed from fragments with
relatively larger numbers of intermediate-mass stars, leading to greater
production of CEMP-$s$ stars.

Simulations of hierarchical galaxy formation suggest that the stars in
the inner halo of the Milky Way formed preferentially from more massive
sub-galactic fragments, capable of supporting extended star-formation
histories \citep[e.g.,][]{Zolotov2009,Font2011,McCarthy2012,
Tissera2013a,Tissera2013b}. By contrast, the outer halo is
predominantly composed of stars formed in lower-mass fragments that were
accreted over long periods of time, experienced short or truncated
star-formation histories, and built up the outer halo as the sum of many
disrupted former satellites. Thus, a possible link exists connecting
higher-mass sub-galactic fragments, steeper IMFs, and the higher
frequency of CEMP-$s$ stars on the one hand, with lower-mass satellites,
flatter IMFs, and a higher frequency of CEMP-no stars on the other.

While these connections are still somewhat speculative, there is now
tentative evidence that lower-mass galaxies may form stars with flatter
IMFs, at least over the limited mass ranges for which star-count IMFs
can be obtained in photometric surveys. In particular, recent HST
studies of the IMF in the Small Magellanic Cloud \citep{Kalirai2013} and
in two Ultra-faint Dwarf (UFD) satellites of the Milky Way
\citep{Brown2012}, indicate a possible flattening trend in the IMF below
$\sim 1$ \msun with declining circular velocity \citep{Geha2013}. This
trend continues for galaxies larger than the Milky Way, which appear to
have formed stars with IMFs steeper than the Salpeter slope
\citep{Conroy2012}. In addition, the limited available high-resolution
spectroscopy for individual stars in UFDs indicates that the majority of
the CEMP stars in these very low-mass galaxies are likely to be of the
CEMP-no sub-class (Frebel \& Norris 2013, Koch et al. 2013, and
reference therein).


This line of reasoning sketches a possible scenario for explaining the
variation of the relative numbers of CEMP-no and CEMP-$s$ associated
with the inner and outer halo. There are several steps in this logic
that are unproven, thus the scenario must still be considered
speculative. For example, the origins of CEMP-no stars may not be
exclusively associated with massive progenitors. The flatter IMFs
observed in the SMC and the UFDs may not continue with that slope into
the range of intermediate and massive stars. Furthermore, the apparent
trend of IMF slope with galaxy mass \citep{Geha2013} may not be strictly
valid, or it may instead be a trend with metallicity or some other
environmental variable. Finally, the frequencies of CEMP stars need to
be reproduced in the context of a detailed model that accounts
simultaneously for the realistic merger histories of the halo components
and the nucleosynthetic origins of their stars.


\acknowledgments

DC is an Australian Research Council Super Science Fellow.
TCB acknowledges partial support from grant PHY 08-22648: Physics
Frontier Center / Joint Institute for Nuclear Astrophysics (JINA),
awarded by the U.S. National Science Foundation.
VMP acknowledges support from the Gemini Observatory.  We thank
an anonymous referee for comments that improved the presentation
in this paper.

This research has made use of the SIMBAD and VIZIER services,
operated at CDS, Strasbourg, France.

\newpage

\newpage

\begin{figure*}
\figurenum{1}
\hspace{2cm}
\includegraphics[width=0.7\textwidth]{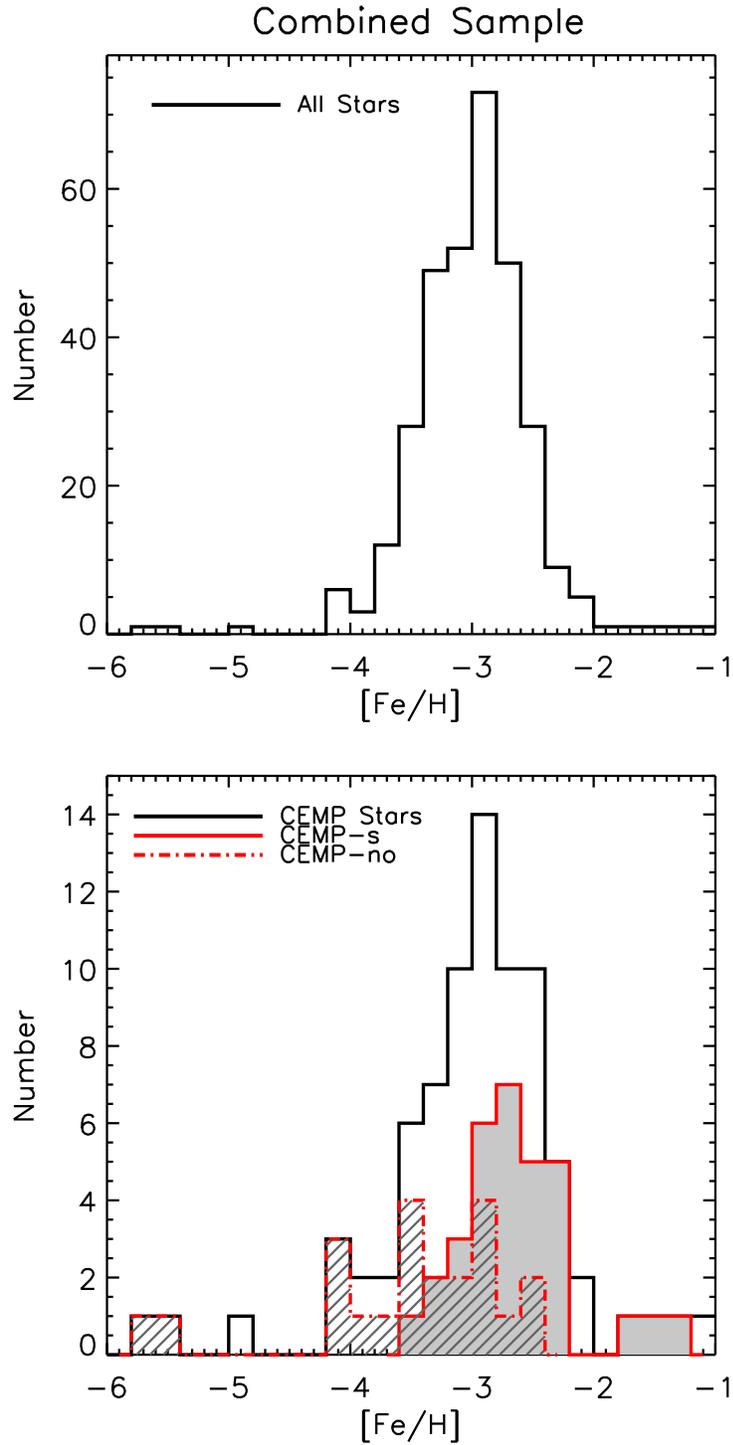}
\caption{Top panel: The metallicity distribution function (MDF) for the
combined sample of stars from \citet{Aoki2013} \& \citet{Yong2013},
323 stars. Bottom panel: MDFs for the
sub-sample of CEMP stars (SS, ES, CSS, or CES; black unshaded
histogram, 96 stars), CEMP-$s$ (SS or ES; gray histogram, red line, 42
stars) and CEMP-no (SS or ES; cross-hatched histogram, dashed red line,
46 stars).}
\end{figure*}

\newpage

\begin{figure*}
\centering
\figurenum{2}
\includegraphics[width=0.6\textwidth]{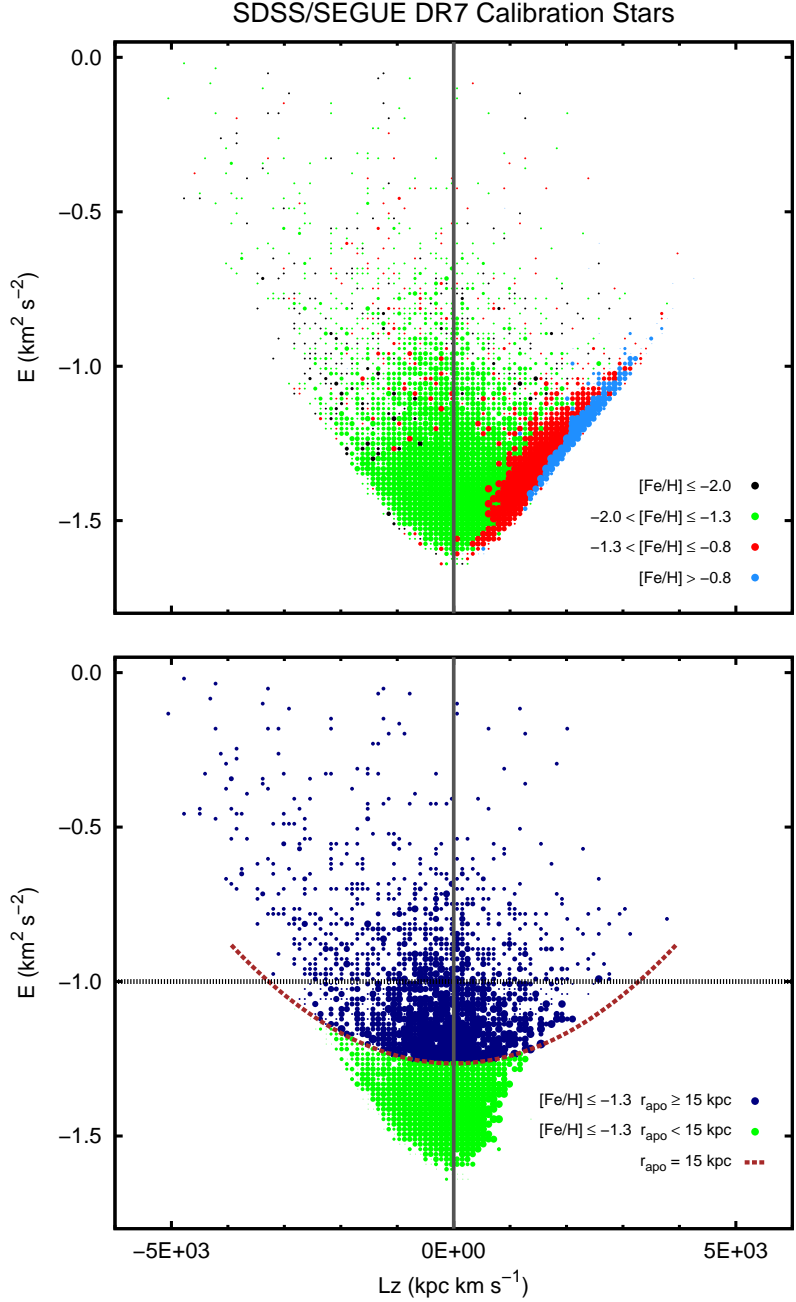}
\caption{Top panel: Lindblad diagram for the sample of 23351 SDSS/SEGUE DR7
calibration stars from C10 (energy is in units of 10$^{5}$). The colors represent
different ranges of metallicity, corresponding to the thick disk (light
blue dots), metal-weak thick disk (red dots), inner halo (green dots),
and outer halo (black dots). The point size is proportional to the
logarithmic number density at a given point of the diagram. The
vertical line divides this diagram at $L_{\rm Z}$ = 0; prograde stars
have $L_{\rm Z}$ $>$ 0, while retrograde stars have $L_{\rm Z}$ $<$ 0.
Bottom panel: Lindblad diagram for the 10431 SDSS/SEGUE DR7 calibration stars with metallicities
[Fe/H] $\le -$1.3. The green dots represent the 7809 stars with apo-Galactic distances
$r_{\rm apo}$ $\le$ 15 kpc, while the blue dots denote the 2622 stars in the same
range of metallicity but with apo-Galactic distance $r_{\rm apo}$ $>$ 15
kpc. The red dashed curve represents the locus of the points with
constant apo-Galactic radii, $r_{\rm apo}$ = 15 kpc, while the
horizontal line at $E = -1.0$ shows (roughly) the energy of the
transition zone between the inner- and outer-halo components (see text).
The vertical line divides this diagram at $L_{\rm Z}$ = 0;
prograde stars have $L_{\rm Z} >$ 0, while retrograde stars have
$L_{\rm Z} <$ 0.  Note the preponderance of retrograde motions
among the likely outer-halo stars located above the $E = -1.0$ line.}
\end{figure*}

\newpage

\begin{figure*}
\figurenum{3}
\hspace{2cm}
\includegraphics[angle=90,width=0.8\textwidth]{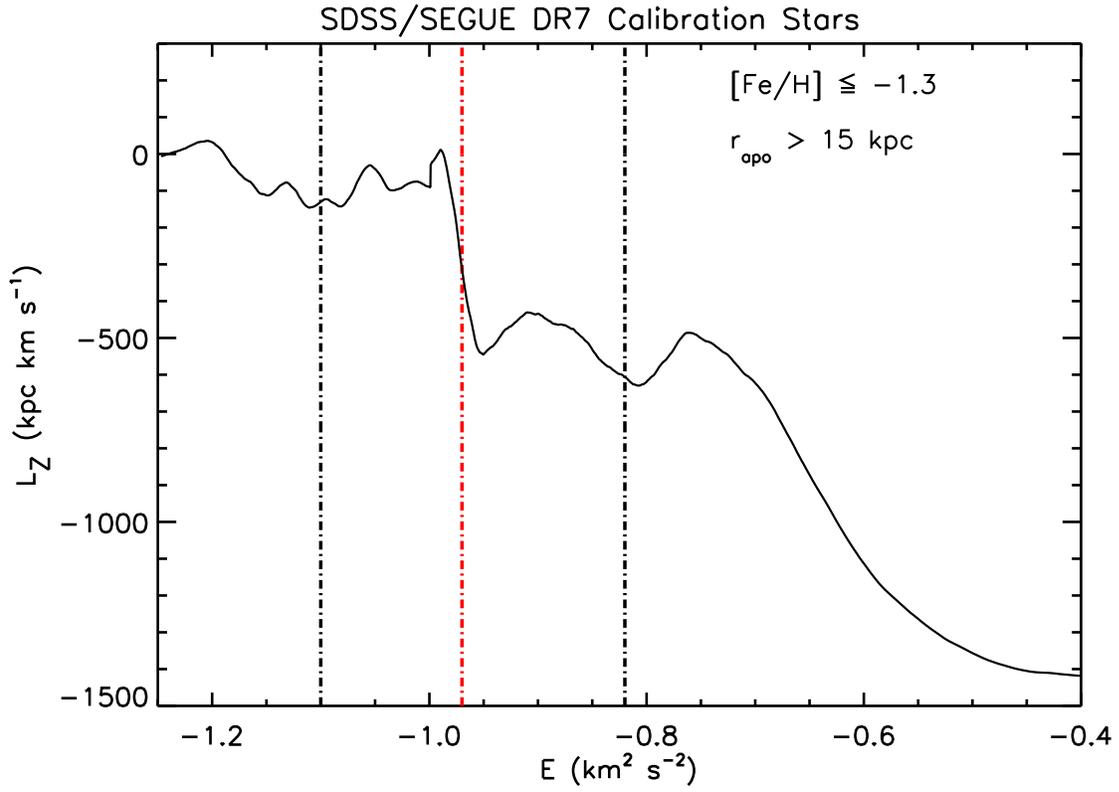}
\caption{Mean angular momentum in the $z$ direction as a function of the
total energy (in units of 10$^{5}$) for the 8039 SDSS/SEGUE DR7 calibration
stars with metallicity [Fe/H] $\le -$1.3 and $r_{\rm apo}$ $>$ 15 kpc. A
moving average with overlapping bins of 200 stars is applied. The
vertical red dot-dashed line denotes the mean point of the transition
zone between the inner halo and outer halo, while the vertical black dot-dashed
lines represents the adopted transition zone.}
\end{figure*}

\newpage

\begin{figure*}
\hspace{-3cm}
\figurenum{4}
\includegraphics[angle=90,width=1.2\textwidth]{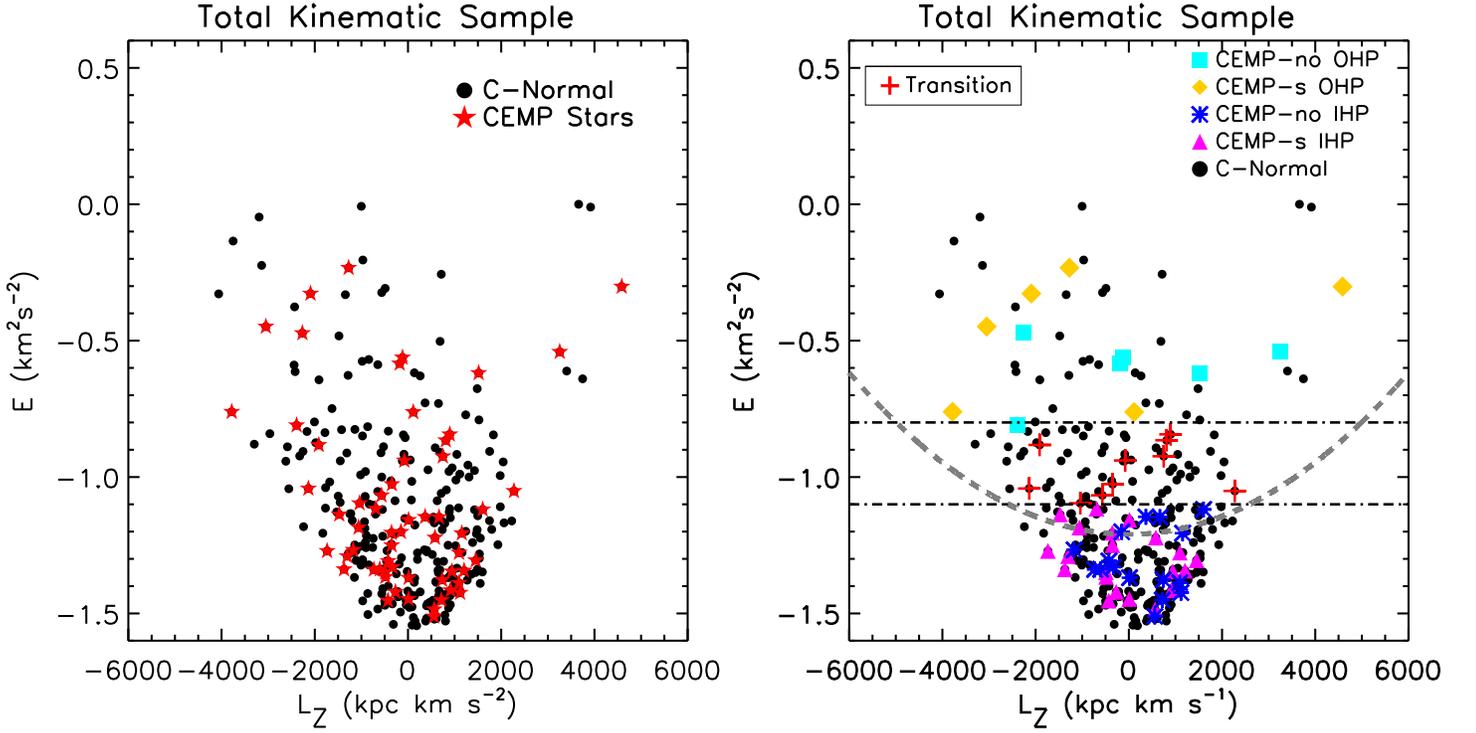}
\caption{Total energy vs. angular momentum in the $z$ direction for the
301 stars in the Total Kinematic Sample. Left panel:
C-normal stars are represented by black dots, while CEMP stars (ES and
CES) are denoted by red stars. Right panel: The same sample of stars, with purple filled
triangles and blue asterisks indicating CEMP-$s$ and CEMP-no members of
the inner halo (ES only), respectively. Yellow filled diamonds
and light blue filled squares represent CEMP-$s$ and CEMP-no members of
the outer halo (ES only), respectively. The red crosses are CEMP-$s$ and
CEMP-no stars located in the transition zone between the inner halo and
the outer halo, with similar probability to be members of these
components (ES only). C-normal (or indeterminate) stars are indicated with black
dots.  The light gray dashed curve denotes the locus of the
points that possess constant apo-Galactic radius, $r_{\rm apo}$ = 15
kpc, while the dark gray dot-dashed horizontal lines shows the values of
the energies delimiting the transition zone. }
\end{figure*}

\newpage
\clearpage
\begin{turnpage}
\begin{deluxetable}{lcrccrcrrrrccrrrccclc}
\tablewidth{0pt}
\tabletypesize{\scriptsize}
\tablenum{1}
\tablecolumns{21}
\tablecaption{Parameters for the Sample of Aoki et al. (2013)}
\tablehead{
\colhead{Name} &
\colhead{~~~} &
\colhead{\teff} &
\colhead{log(g)} &
\colhead{[Fe/H]} &
\colhead{V$_{rad}$} &
\colhead{$\sigma_{V_{rad}}$} &
\colhead{RA} &
\colhead{DEC} &
\colhead{PM$_{RA}$} &
\colhead{PM$_{DEC}$} &
\colhead{$\sigma_{PM_{RA}}$} &
\colhead{$\sigma_{PM_{DEC}}$} &
\colhead{d} &
\colhead{[C/Fe]} &
\colhead{[Ba/Fe]} &
\colhead{C} &
\colhead{Ba} &
\colhead{CEMP} &
\colhead{Sample} &
\colhead{Member} \\
\colhead{} &
\colhead{} &
\colhead{(K)} &
\colhead{} &
\colhead{} &
\colhead{(km~s$^{-1}$)} &
\colhead{(km~s$^{-1}$)} &
\colhead{(deg)} &
\colhead{(deg)} &
\colhead{(mas~yr$^{-1}$)} &
\colhead{(mas~yr$^{-1}$)} &
\colhead{(mas~yr$^{-1}$)} &
\colhead{(mas~yr$^{-1}$)} &
\colhead{(kpc)} &
\colhead{} &
\colhead{} &
\colhead{} &
\colhead{} &
\colhead{Class} &
\colhead{} &
\colhead{} \\
\colhead{(1)} &
\colhead{} &
\colhead{(2)} &
\colhead{(3)} &
\colhead{(4)} &
\colhead{(5)} &
\colhead{(6)} &
\colhead{(7)} &
\colhead{(8)} &
\colhead{(9)} &
\colhead{(10)} &
\colhead{(11)} &
\colhead{(12)} &
\colhead{(13)} &
\colhead{(14)} &
\colhead{(15)} &
\colhead{(16)} &
\colhead{(17)} &
\colhead{(18)} &
\colhead{(19)} &
\colhead{(20)}
}
\startdata
2803-54368-459                  &  &   6150    &         4.0    &       $-$3.26    &      $-$308.2  &         0.2  &       0.583  &        29.481  &          0.15  &      $-$21.13  &        2.50  &        2.50  &        1.55   &    2.63  &   1.84  &  3  & 3 & s  & 1100 & I \\
2624-54380-061                  &  &   4950    &         2.0    &       $-$2.66    &         117.2  &         0.1  &       2.052  &      $-$5.491  &       $-$0.42  &          0.91  &        2.42  &        2.42  &       31.03   & \nodata  &   0.83  &  0  & 3 &    &         & O \\
1912-53293-352                  &  &   4600    &         5.0    &       $-$2.65    &      $-$122.9  &         0.1  &       4.585  &      $-$9.661  &          2.62  &     $-$109.50  &        2.78  &        2.78  &        0.47   & $-$0.88  & \nodata &  2  & 0 &    &         & I \\
1121-52873-136                  &  &   6450    &         4.0    &       $-$2.81    &           3.3  &         0.1  &       5.064  &      $-$0.683  &          8.62  &          1.74  &        2.56  &        2.56  &        3.45   & \nodata  & \nodata &  0  & 0 &    &         & I \\
0390-51816-187                  &  &   6650    &         4.0    &       $-$3.16    &       $-$94.6  &         0.2  &       5.307  &      $-$0.835  &          8.45  &       $-$0.74  &        2.49  &        2.49  &        3.17   & \nodata  & \nodata &  0  & 0 &    &         & I \\
0688-52203-152                  &  &   5500    &         4.0    &       $-$3.12    &          89.8  &         0.1  &       5.984  &      $-$0.053  &         14.47  &       $-$6.00  &        2.60  &        2.60  &        2.56   & \nodata  & $-$0.42 &  0  & 4 &    &         & I \\
0753-52233-013                  &  &   6250    &         4.0    &       $-$3.03    &          27.7  &         0.2  &       6.956  &        14.072  &         13.87  &       $-$2.83  &        2.75  &        2.75  &        3.06   & \nodata  & \nodata &  0  & 0 &    &         & I \\
\end{deluxetable}
\end{turnpage}

\clearpage
\begin{turnpage}
\begin{deluxetable}{lcrccrcrrrrccrrrccclc}
\tablewidth{0pt}
\tabletypesize{\scriptsize}
\tablenum{1}
\tablecolumns{21}
\tablecaption{Parameters for the Sample of Aoki et al. (2013)}
\tablehead{
\colhead{Name} &
\colhead{~~~} &
\colhead{\teff} &
\colhead{log(g)} &
\colhead{[Fe/H]} &
\colhead{V$_{rad}$} &
\colhead{$\sigma_{V_{rad}}$} &
\colhead{RA} &
\colhead{DEC} &
\colhead{PM$_{RA}$} &
\colhead{PM$_{DEC}$} &
\colhead{$\sigma_{PM_{RA}}$} &
\colhead{$\sigma_{PM_{DEC}}$} &
\colhead{d} &
\colhead{[C/Fe]} &
\colhead{[Ba/Fe]} &
\colhead{C} &
\colhead{Ba} &
\colhead{CEMP} &
\colhead{Sample} &
\colhead{Member} \\
\colhead{} &
\colhead{} &
\colhead{(K)} &
\colhead{} &
\colhead{} &
\colhead{(km~s$^{-1}$)} &
\colhead{(km~s$^{-1}$)} &
\colhead{(deg)} &
\colhead{(deg)} &
\colhead{(mas~yr$^{-1}$)} &
\colhead{(mas~yr$^{-1}$)} &
\colhead{(mas~yr$^{-1}$)} &
\colhead{(mas~yr$^{-1}$)} &
\colhead{(kpc)} &
\colhead{} &
\colhead{} &
\colhead{} &
\colhead{} &
\colhead{Class} &
\colhead{} &
\colhead{} \\
\colhead{(1)} &
\colhead{} &
\colhead{(2)} &
\colhead{(3)} &
\colhead{(4)} &
\colhead{(5)} &
\colhead{(6)} &
\colhead{(7)} &
\colhead{(8)} &
\colhead{(9)} &
\colhead{(10)} &
\colhead{(11)} &
\colhead{(12)} &
\colhead{(13)} &
\colhead{(14)} &
\colhead{(15)} &
\colhead{(16)} &
\colhead{(17)} &
\colhead{(18)} &
\colhead{(19)} &
\colhead{(20)}
}
\startdata
2387-53770-316                  &  &   6100    &         4.0    &       $-$3.09    &       $-$57.9  &         0.1  &     151.115  &        34.713  &       $-$2.92  &      $-$15.62  &        2.57  &        2.57  &        1.34   & \nodata  & \nodata &  0  & 0 &    &         & I \\
1430-53002-498                  &  &   6600    &         4.0    &       $-$3.06    &      $-$133.1  &         0.2  &     158.256  &        40.018  &          3.54  &         10.76  &        2.55  &        2.55  &        2.73   & \nodata  & \nodata &  0  & 0 &    &         & O \\
1600-53090-378                  &  &   5850    &         4.0    &       $-$3.47    &       $-$33.5  &         0.1  &     159.208  &        12.205  &          3.74  &      $-$18.69  &        2.47  &        2.47  &        1.85   & $+$1.84  & $+$1.35 &  3  & 3 & s  & 1100 & I \\
0581-52356-245                  &  &   6300    &         4.0    &       $-$2.88    &         159.8  &         0.1  &     166.544  &         3.723  &       $-$2.83  &      $-$35.60  &        2.67  &        2.67  &        2.97   & \nodata  & \nodata &  0  & 0 &    &         & T \\
2491-53855-389                  &  &   6050    &         4.0    &       $-$3.17    &       $-$27.9  &         0.6  &     167.090  &        17.796  &      $-$11.99  &      $-$40.35  &        2.38  &        2.38  &        2.00   & \nodata  & \nodata &  0  & 0 &    &         & I \\
1979-53431-181                  &  &   5750    &         4.0    &       $-$3.14    &          47.5  &         0.1  &     170.216  &        30.457  &      $-$20.14  &      $-$44.00  &        2.34  &        2.34  &        2.07   & \nodata  & \nodata &  0  & 0 &    &         & I \\
2036-53446-324\tablenotemark{f} &  &   6550    &         4.0    &       $-$2.82    &       $-$15.9  &         0.4  &     172.057  &        38.697  &      $-$25.73  &       $-$1.71  &        2.49  &        2.49  &        2.12   & \nodata  & \nodata &  0  & 0 &    &         & U \\
\end{deluxetable}
\end{turnpage}

\clearpage
\begin{turnpage}
\begin{deluxetable}{lcrccrcrrrrccrrrccclc}
\tablewidth{0pt}
\tabletypesize{\scriptsize}
\tablenum{1}
\tablecolumns{21}
\tablecaption{Parameters for the Sample of Aoki et al. (2013)}
\tablehead{
\colhead{Name} &
\colhead{~~~} &
\colhead{\teff} &
\colhead{log(g)} &
\colhead{[Fe/H]} &
\colhead{V$_{rad}$} &
\colhead{$\sigma_{V_{rad}}$} &
\colhead{RA} &
\colhead{DEC} &
\colhead{PM$_{RA}$} &
\colhead{PM$_{DEC}$} &
\colhead{$\sigma_{PM_{RA}}$} &
\colhead{$\sigma_{PM_{DEC}}$} &
\colhead{d} &
\colhead{[C/Fe]} &
\colhead{[Ba/Fe]} &
\colhead{C} &
\colhead{Ba} &
\colhead{CEMP} &
\colhead{Sample} &
\colhead{Member} \\
\colhead{} &
\colhead{} &
\colhead{(K)} &
\colhead{} &
\colhead{} &
\colhead{(km~s$^{-1}$)} &
\colhead{(km~s$^{-1}$)} &
\colhead{(deg)} &
\colhead{(deg)} &
\colhead{(mas~yr$^{-1}$)} &
\colhead{(mas~yr$^{-1}$)} &
\colhead{(mas~yr$^{-1}$)} &
\colhead{(mas~yr$^{-1}$)} &
\colhead{(kpc)} &
\colhead{} &
\colhead{} &
\colhead{} &
\colhead{} &
\colhead{Class} &
\colhead{} &
\colhead{} \\
\colhead{(1)} &
\colhead{} &
\colhead{(2)} &
\colhead{(3)} &
\colhead{(4)} &
\colhead{(5)} &
\colhead{(6)} &
\colhead{(7)} &
\colhead{(8)} &
\colhead{(9)} &
\colhead{(10)} &
\colhead{(11)} &
\colhead{(12)} &
\colhead{(13)} &
\colhead{(14)} &
\colhead{(15)} &
\colhead{(16)} &
\colhead{(17)} &
\colhead{(18)} &
\colhead{(19)} &
\colhead{(20)}
}
\startdata
2174-53521-423                  &  &   6450    &         4.0    &       $-$3.39    &       $-$51.1  &         0.3  &     250.022  &        37.152  &       $-$5.72  &          5.64  &        2.72  &        2.72  &        2.10   & \nodata  & \nodata &  0  & 0 &    &         & I \\
1690-53475-323\tablenotemark{c} &  &   6100    &         4.0    &       $-$3.05    &       $-$18.7  &         0.2  &     251.542  &        28.406  &      $-$16.01  &         45.77  &        2.59  &        2.59  &        2.07   & $+$2.52  & $+$1.78 &  3  & 3 & s  & 1100 & U \\
2180-54613-258                  &  &   6600    &         4.0    &       $-$2.56    &       $-$33.1  &         0.3  &     252.569  &        22.704  &       $-$5.93  &       $-$8.72  &        2.68  &        2.68  &        2.68   & \nodata  & \nodata &  0  & 0 &    &         & I \\
0974-52427-332                  &  &   6050    &         4.0    &       $-$3.24    &       $-$42.5  &         0.1  &     254.895  &        35.265  &      $-$17.92  &      $-$37.19  &        2.73  &        2.73  &        3.18   & \nodata  & \nodata &  0  & 0 &    &         & O \\
2808-54524-510                  &  &   5100    &         4.8    &       $-$3.21    &      $-$174.9  &         0.2  &     255.915  &        28.614  &      $-$27.48  &     $-$129.90  &        2.46  &        2.46  &        0.57   & $+$0.28  & \nodata &  2  & 0 &    &         & I \\
2797-54616-258                  &  &   6350    &         4.0    &       $-$2.85    &      $-$353.4  &         0.2  &     262.195  &         6.951  &          2.04  &      $-$19.18  &        2.28  &        2.28  &        2.81   & \nodata  & \nodata &  0  & 0 &    &         & I \\
2799-54368-138                  &  &   5200    &         2.7    &       $-$2.51    &      $-$211.9  &         0.1  &     263.575  &        43.268  &          0.79  &       $-$5.95  &        2.45  &        2.45  &       15.86   & $+$1.78  & $+$1.61 &  3  & 3 & s  & 1100 & O \\
\enddata
\end{deluxetable}
\end{turnpage}

\clearpage
\begin{turnpage}
\begin{deluxetable}{lcrccrcrrrrccrrrccclc}
\tablewidth{0pt}
\tabletypesize{\scriptsize}
\tablenum{2}
\tablecolumns{21}
\tablecaption{Parameters for the Sample of Yong et al. (2013)}
\tablehead{
\colhead{Name} &
\colhead{~~~} &
\colhead{\teff} &
\colhead{log(g)} &
\colhead{[Fe/H]} &
\colhead{V$_{rad}$} &
\colhead{$\sigma_{V_{rad}}$} &
\colhead{RA} &
\colhead{DEC} &
\colhead{PM$_{RA}$} &
\colhead{PM$_{DEC}$} &
\colhead{$\sigma_{PM_{RA}}$} &
\colhead{$\sigma_{PM_{DEC}}$} &
\colhead{d} &
\colhead{[C/Fe]} &
\colhead{[Ba/Fe]} &
\colhead{C} &
\colhead{Ba} &
\colhead{CEMP} &
\colhead{Sample} &
\colhead{Member} \\
\colhead{} &
\colhead{} &
\colhead{(K)} &
\colhead{} &
\colhead{} &
\colhead{(km~s$^{-1}$)} &
\colhead{(km~s$^{-1}$)} &
\colhead{(deg)} &
\colhead{(deg)} &
\colhead{(mas~yr$^{-1}$)} &
\colhead{(mas~yr$^{-1}$)} &
\colhead{(mas~yr$^{-1}$)} &
\colhead{(mas~yr$^{-1}$)} &
\colhead{(kpc)} &
\colhead{} &
\colhead{} &
\colhead{} &
\colhead{} &
\colhead{Class} &
\colhead{} &
\colhead{} \\
\colhead{(1)} &
\colhead{} &
\colhead{(2)} &
\colhead{(3)} &
\colhead{(4)} &
\colhead{(5)} &
\colhead{(6)} &
\colhead{(7)} &
\colhead{(8)} &
\colhead{(9)} &
\colhead{(10)} &
\colhead{(11)} &
\colhead{(12)} &
\colhead{(13)} &
\colhead{(14)} &
\colhead{(15)} &
\colhead{(16)} &
\colhead{(17)} &
\colhead{(18)} &
\colhead{(19)} &
\colhead{(20)}
}
\startdata
CS 22957-022                     &  &   5146    &         2.4    &       $-$2.92    &       $-$33.8  &         0.2  &       0.440  &      $-$5.830  &           1.1  &       $-$28.4  &        2.7  &        2.7  &        3.30    & $+$0.16   & $-$1.09   &  2 &  4 &     &         & T \\
CS 29503-010\tablenotemark{a}    &  &   6570    &         4.3    &       $-$1.00    &       $-$24.3  &         0.1  &       1.231  &     $-$24.405  &       $-$11.5  &        $-$3.6  &        1.9  &        1.6  &     \nodata    & $+$1.07   & $+$1.51   &  3 &  3 &  s  & 1100 & U \\
CS 31085-024                     &  &   5778    &         4.6    &       $-$2.80    &      $-$322.0  &         0.2  &       2.116  &        10.906  &          66.0  &       $-$36.0  &        2.0  &        2.0  &        0.38    & $+$0.36   & $-$0.57   &  2 &  4 &     &         & I \\
BS 17570-063                     &  &   6233    &         4.5    &       $-$2.95    &         184.4  &         0.3  &       5.151  &        23.794  &          38.0  &       $-$30.0  &        4.0  &        1.0  &        0.77    & $+$0.40   & $-$0.26   &  2 &  4 &     &         & T \\
HE 0024-2523                     &  &   6635    &         4.1    &       $-$2.82    &      $-$181.6  &         1.0  &       6.865  &     $-$25.108  &       $-$23.7  &       $-$66.4  &        2.0  &        2.2  &        1.75    & $+$2.62   & $+$1.52   &  3 &  3 &  s  & 1100 & O \\
CS 29527-015\tablenotemark{j}    &  &   6577    &         3.9    &       $-$3.32    &          50.9  &         0.3  &       7.295  &     $-$19.169  &          64.0  &       $-$33.8  &        1.9  &        1.9  &        1.62    & $+$1.18   & \nodata   &  3 &  0 &     & 0011 & O \\
CS 30339-069                     &  &   6326    &         3.8    &       $-$3.05    &          34.8  &         0.3  &       7.567  &     $-$35.948  &          25.7  &       $-$23.8  &        3.5  &        2.0  &        2.12    & $+$0.56   & \nodata   &  2 &  0 &     & 0001 & I \\
\end{deluxetable}
\end{turnpage}

\clearpage

\begin{turnpage}
\begin{deluxetable}{lcrccrcrrrrccrrrccclc}
\tablewidth{0pt}
\tabletypesize{\scriptsize}
\tablenum{2}
\tablecolumns{21}
\tablecaption{Parameters for the Sample of Yong et al. (2013)}
\tablehead{
\colhead{Name} &
\colhead{~~~} &
\colhead{\teff} &
\colhead{log(g)} &
\colhead{[Fe/H]} &
\colhead{V$_{rad}$} &
\colhead{$\sigma_{V_{rad}}$} &
\colhead{RA} &
\colhead{DEC} &
\colhead{PM$_{RA}$} &
\colhead{PM$_{DEC}$} &
\colhead{$\sigma_{PM_{RA}}$} &
\colhead{$\sigma_{PM_{DEC}}$} &
\colhead{d} &
\colhead{[C/Fe]} &
\colhead{[Ba/Fe]} &
\colhead{C} &
\colhead{Ba} &
\colhead{CEMP} &
\colhead{Sample} &
\colhead{Member} \\
\colhead{} &
\colhead{} &
\colhead{(K)} &
\colhead{} &
\colhead{} &
\colhead{(km~s$^{-1}$)} &
\colhead{(km~s$^{-1}$)} &
\colhead{(deg)} &
\colhead{(deg)} &
\colhead{(mas~yr$^{-1}$)} &
\colhead{(mas~yr$^{-1}$)} &
\colhead{(mas~yr$^{-1}$)} &
\colhead{(mas~yr$^{-1}$)} &
\colhead{(kpc)} &
\colhead{} &
\colhead{} &
\colhead{} &
\colhead{} &
\colhead{Class} &
\colhead{} &
\colhead{} \\
\colhead{(1)} &
\colhead{} &
\colhead{(2)} &
\colhead{(3)} &
\colhead{(4)} &
\colhead{(5)} &
\colhead{(6)} &
\colhead{(7)} &
\colhead{(8)} &
\colhead{(9)} &
\colhead{(10)} &
\colhead{(11)} &
\colhead{(12)} &
\colhead{(13)} &
\colhead{(14)} &
\colhead{(15)} &
\colhead{(16)} &
\colhead{(17)} &
\colhead{(18)} &
\colhead{(19)} &
\colhead{(20)}
}
\startdata
CS 22172-002                     &  &   4893    &         1.7    &       $-$3.77    &         251.0  &         1.0  &      48.587  &     $-$10.586  &           7.0  &       $-$14.3  &        1.4  &        2.0  &        4.82    &    0.00   & $-$1.17   &  2 &  4 &     &         & T \\
HE 0314-1739                     &  &   6570    &         4.3    &       $-$2.86    &          40.9  &         0.5  &      49.257  &     $-$17.482  &          10.0  &       $-$26.5  &        4.2  &        4.2  &        2.59    & $< +$1.49 & $< -$0.27 &  1 &  4 &     &         & I \\
HE 0336+0113\tablenotemark{c}    &  &   5819    &         3.6    &       $-$2.60    &          66.6  &         1.5  &      54.720  &         1.385  &         106.7  &       $-$38.7  &        7.7  &        7.7  &        2.11    & $+$2.25   & $+$2.69   &  3 &  3 &  s  & 1100 & U \\
HE 0355-3728                     &  &   6418    &         4.1    &       $-$3.41    &          82.0  &         1.8  &      59.152  &     $-$44.568  &           1.2  &       $-$29.1  &        5.2  &        5.2  &        1.80    & $< +$2.14 & $< +$0.21 &  1 &  1 &     &         & I \\
HE 0400-2030                     &  &   5579    &         3.5    &       $-$1.68    &         221.8  &         0.1  &      60.562  &     $-$20.365  &           3.6  &        $-$4.4  &        1.8  &        1.8  &        1.28    & $+$1.14   & $+$1.85   &  3 &  3 &  s  & 1100 & I \\
CS 22177-009                     &  &   6423    &         4.4    &       $-$2.98    &      $-$208.4  &         0.3  &      61.919  &     $-$25.046  &          20.5  &       $-$60.7  &        1.4  &        1.4  &        0.84    & $+$0.38   & \nodata   &  2 &  0 &     &         & T \\
CS 22169-035                     &  &   4654    &         1.1    &       $-$2.95    &          16.1  &         0.9  &      63.058  &     $-$12.085  &           7.8  &        $-$6.8  &        2.0  &        2.1  &        9.39    & $-$0.24   & $-$1.19   &  2 &  4 &     &         & T \\
\end{deluxetable}
\end{turnpage}

\clearpage

\begin{turnpage}
\begin{deluxetable}{lcrccrcrrrrccrrrccclc}
\tablewidth{0pt}
\tabletypesize{\scriptsize}
\tablenum{2}
\tablecolumns{21}
\tablecaption{Parameters for the Sample of Yong et al. (2013)}
\tablehead{
\colhead{Name} &
\colhead{~~~} &
\colhead{\teff} &
\colhead{log(g)} &
\colhead{[Fe/H]} &
\colhead{V$_{rad}$} &
\colhead{$\sigma_{V_{rad}}$} &
\colhead{RA} &
\colhead{DEC} &
\colhead{PM$_{RA}$} &
\colhead{PM$_{DEC}$} &
\colhead{$\sigma_{PM_{RA}}$} &
\colhead{$\sigma_{PM_{DEC}}$} &
\colhead{d} &
\colhead{[C/Fe]} &
\colhead{[Ba/Fe]} &
\colhead{C} &
\colhead{Ba} &
\colhead{CEMP} &
\colhead{Sample} &
\colhead{Member} \\
\colhead{} &
\colhead{} &
\colhead{(K)} &
\colhead{} &
\colhead{} &
\colhead{(km~s$^{-1}$)} &
\colhead{(km~s$^{-1}$)} &
\colhead{(deg)} &
\colhead{(deg)} &
\colhead{(mas~yr$^{-1}$)} &
\colhead{(mas~yr$^{-1}$)} &
\colhead{(mas~yr$^{-1}$)} &
\colhead{(mas~yr$^{-1}$)} &
\colhead{(kpc)} &
\colhead{} &
\colhead{} &
\colhead{} &
\colhead{} &
\colhead{Class} &
\colhead{} &
\colhead{} \\
\colhead{(1)} &
\colhead{} &
\colhead{(2)} &
\colhead{(3)} &
\colhead{(4)} &
\colhead{(5)} &
\colhead{(6)} &
\colhead{(7)} &
\colhead{(8)} &
\colhead{(9)} &
\colhead{(10)} &
\colhead{(11)} &
\colhead{(12)} &
\colhead{(13)} &
\colhead{(14)} &
\colhead{(15)} &
\colhead{(16)} &
\colhead{(17)} &
\colhead{(18)} &
\colhead{(19)} &
\colhead{(20)}
}
\startdata
HE 1424-0241                     &  &   5260    &         2.7    &       $-$4.05    &          60.4  &         1.5  &     216.668  &      $-$2.908  &        $-$2.8  &        $-$1.2  &        4.0  &        4.0  &        5.29    & $< +$0.63 & $-$0.75   &  2 &  4 &     &         & I \\
HD 126587                        &  &   4853    &         1.6    &       $-$3.00    &         148.7  &         0.7  &     216.752  &     $-$22.244  &       $-$14.4  &       $-$51.3  &        0.8  &        0.8  &        0.93    & $+$0.19   & $-$0.10   &  2 &  4 &     &         & I \\
HE 1429-0551                     &  &   4757    &         1.4    &       $-$2.60    &       $-$44.9  &         0.1  &     218.130  &      $-$6.083  &       $-$15.9  &       $-$19.5  &        4.0  &        4.0  &        6.06    & $+$2.28   & $+$1.47   &  3 &  3 &  s  & 1100 & O \\
BS 16477-003                     &  &   4879    &         1.7    &       $-$3.39    &      $-$222.6  &         1.0  &     218.237  &         6.769  &       $-$10.1  &        $-$4.0  &        4.0  &        4.0  &       10.41    & $+$0.29   & $-$0.45   &  2 &  4 &     &         & O \\
CS 30325-094                     &  &   4948    &         1.9    &       $-$3.35    &      $-$157.8  &         1.0  &     223.664  &         4.361  &       $-$32.2  &       $-$20.1  &        2.0  &        2.0  &        3.57    &    0.00   & $-$1.88   &  2 &  4 &     &         & O \\
BS 16968-061                     &  &   6343    &         3.8    &       $-$2.85    &       $-$80.7  &         0.3  &     225.275  &         3.713  &       $-$95.4  &       $-$89.5  &        9.0  &        9.0  &        0.98    & $+$0.45   & \nodata   &  2 &  0 &     &         & T \\
CS 30325-028                     &  &   4894    &         1.7    &       $-$2.87    &         148.8  &         0.2  &     226.333  &         6.285  &       $-$10.1  &           4.9  &        2.4  &        2.4  &        5.03    & $+$0.38   & $-$0.49   &  2 &  4 &     &         & T \\
\end{deluxetable}
\end{turnpage}

\clearpage
\begin{turnpage}
\begin{deluxetable}{lcrccrcrrrrccrrrccclc}
\tablewidth{0pt}
\tabletypesize{\scriptsize}
\tablenum{2}
\tablecolumns{21}
\tablecaption{Parameters for the Sample of Yong et al. (2013)}
\tablehead{
\colhead{Name} &
\colhead{~~~} &
\colhead{\teff} &
\colhead{log(g)} &
\colhead{[Fe/H]} &
\colhead{V$_{rad}$} &
\colhead{$\sigma_{V_{rad}}$} &
\colhead{RA} &
\colhead{DEC} &
\colhead{PM$_{RA}$} &
\colhead{PM$_{DEC}$} &
\colhead{$\sigma_{PM_{RA}}$} &
\colhead{$\sigma_{PM_{DEC}}$} &
\colhead{d} &
\colhead{[C/Fe]} &
\colhead{[Ba/Fe]} &
\colhead{C} &
\colhead{Ba} &
\colhead{CEMP} &
\colhead{Sample} &
\colhead{Member} \\
\colhead{} &
\colhead{} &
\colhead{(K)} &
\colhead{} &
\colhead{} &
\colhead{(km~s$^{-1}$)} &
\colhead{(km~s$^{-1}$)} &
\colhead{(deg)} &
\colhead{(deg)} &
\colhead{(mas~yr$^{-1}$)} &
\colhead{(mas~yr$^{-1}$)} &
\colhead{(mas~yr$^{-1}$)} &
\colhead{(mas~yr$^{-1}$)} &
\colhead{(kpc)} &
\colhead{} &
\colhead{} &
\colhead{} &
\colhead{} &
\colhead{Class} &
\colhead{} &
\colhead{} \\
\colhead{(1)} &
\colhead{} &
\colhead{(2)} &
\colhead{(3)} &
\colhead{(4)} &
\colhead{(5)} &
\colhead{(6)} &
\colhead{(7)} &
\colhead{(8)} &
\colhead{(9)} &
\colhead{(10)} &
\colhead{(11)} &
\colhead{(12)} &
\colhead{(13)} &
\colhead{(14)} &
\colhead{(15)} &
\colhead{(16)} &
\colhead{(17)} &
\colhead{(18)} &
\colhead{(19)} &
\colhead{(20)}
}
\startdata
CS 22948-066\tablenotemark{i}    &  &   5077    &         2.2    &       $-$3.20    &      $-$170.6  &         1.0  &     326.213  &     $-$37.465  &        $-$2.0  &        $-$5.5  &        1.2  &        1.2  &        4.42    & $-$0.74   & $-$1.15   &  2 &  4 &     &         & I \\
HE 2142-5656                     &  &   4939    &         1.9    &       $-$2.87    &         103.4  &         0.5  &     326.585  &     $-$56.705  &        $-$7.0  &        $-$5.2  &        0.7  &        0.8  &        6.86    & $+$0.95   & $-$0.63   &  3 &  4 &  no & 1100 & I \\
CS 22944-032                     &  &   5293    &         2.8    &       $-$2.98    &          17.1  &         0.2  &     326.930  &     $-$13.673  &       $-$10.2  &       $-$47.0  &        1.8  &        1.9  &        2.09    & $+$0.31   & $-$0.76   &  2 &  4 &     &         & T \\
CS 22948-093                     &  &   6609    &         4.2    &       $-$3.17    &         364.3  &         0.3  &     327.631  &     $-$41.130  &       $-$18.4  &       $-$20.3  &        2.1  &        1.9  &        1.69    & \nodata   & \nodata   &  0 &  0 &     &         & O \\
CS 22956-050                     &  &   4844    &         1.6    &       $-$3.39    &        $-$0.1  &         1.0  &     329.524  &     $-$65.224  &        $-$2.4  &        $-$3.1  &        2.1  &        2.2  &       11.91    & $+$0.27   & $-$0.78   &  2 &  4 &     &         & I \\
HE 2158-0348\tablenotemark{c}    &  &   5150    &         2.4    &       $-$2.57    &          67.6  &         1.5  &     330.167  &      $-$3.570  &          12.4  &       $-$11.1  &        3.8  &        3.8  &        8.83    & $+$1.87   & $+$1.75   &  3 &  3 &  s  & 1100 & U \\
BS 17569-049                     &  &   4645    &         1.1    &       $-$2.84    &      $-$213.1  &         1.0  &     331.243  &         4.026  &           0.6  &        $-$4.4  &        3.8  &        3.8  &       11.61    & $-$0.12   & $+$0.20   &  2 &  2 &     &         & I \\
\enddata
\end{deluxetable}
\end{turnpage}

\clearpage
\begin{deluxetable*}{cccc}
\tablenum{3}
\tablewidth{0pc}
\tablecaption{Kinematic Membership Summary \label{tabkin}}
\tablehead{
\colhead{Member}       &
\colhead{Total}        &
\colhead{~~~~~~ no ~~~~~~}           &
\colhead{~~~~~~ s ~~~~~~}            \\
}

\startdata
\multicolumn{4}{c}{Total Kinematic Sample}          \\
\hline
  O &  46 & \nodata & \nodata \\
  I & 190 & \nodata & \nodata \\
  T &  65 & \nodata & \nodata \\
  U &  22 & \nodata & \nodata \\
\hline
\multicolumn{4}{c}{CEMP Kinematic Sample}             \\
\hline
  O &  12 (20) &   6 (14) &   6  (6) \\
  I &  37 (48) &  16 (27) &  21 (21) \\
  T &   5 (10) &   0  (3) &   5  (7) \\
  U &  10 (10) &   2  (2) &   8  (8)
\enddata
\tablecomments{The column labeled ``Member'' indicates the halo component
into which stars with kinematic information have been assigned,
based on the criteria described in the text.  ``O'' indicates outer halo,
``I'' indicates inner halo, ``T'' indicates the transition zone, and ``U''
indicates that the star was unclassified (unbound orbits).  The upper section
of the table pertains to the Total Kinematic Sample of program stars, including those
without measurements of carbon available, or that are simply C-normal.  The lower
section pertains to the CEMP Kinematic Sample, including only stars that are
considered C-enhanced.  The first number in each column pertains
to stars in the SS, the second number (in parentheses) in each column pertains
to stars in the ES.}
\end{deluxetable*}

\end{document}